%% file: main.tex
\documentclass[journal=jacsat,manuscript=article]{achemso}

\usepackage{chemformula} 
\usepackage[T1]{fontenc} 
\usepackage[utf8]{inputenc}
\usepackage{comment}

\usepackage{tabularx}
\usepackage{siunitx}
\usepackage{tikz}
\usepackage{subcaption}
\usetikzlibrary{positioning}
\usetikzlibrary{calc}
\usepackage{hyperref}
\usepackage{amssymb}

\def \Ss{\mathbf{\tilde S}_s}
\def \SS{\mathbf{\tilde S}_s^2}
\def \SSS{\mathbf{\tilde S}_s^3}

\def \Rs{\mathbf{\tilde R}_s}
\def \RR{\mathbf{\tilde R}_s^2}

\def\tr{\mathrm{tr}}

\usepackage{soul,color}
\usepackage[normalem]{ulem}
\usepackage{relsize}

\include{symbols_and_cmd}

\author{Baptiste Hardy}
\email{b.hardy@tudelft.nl}
\affiliation[TUDelft]
{Process and Energy Department, TU Delft, Delft, Netherlands}

\author{Stefanie Rauchenzauner}
\affiliation[TUMunich]{Chair of Process Systems Engineering, Technical University of Munich, Freising, Germany}

\author{Pascal Fede}
\affiliation[IMFT]
{Institut de Mécanique des Fluides de Toulouse, Université de Toulouse, CNRS, France}

\author{Simon Schneiderbauer}
\affiliation[JKU]{Department of Particulate Flow Modelling, Johannes Kepler University, Linz, Austria}

\author{Olivier Simonin}
\affiliation[IMFT]
{Institut de Mécanique des Fluides de Toulouse, Université de Toulouse, CNRS, France}

\author{Sankaran Sundaresan}
\affiliation[Princenton]
{Department of Chemical and Biological Engineering Princeton University, Princeton, USA}

\author{Ali Ozel}
\affiliation[HeriotWatt]
{School of Engineering and Physical Sciences, Heriot-Watt University, Edinburgh, EH14 4AS, UK}

\title[An \textsf{achemso} dem]{Machine learning approaches to close the filtered two-fluid model for gas-solid flows: Models for subgrid drag force and solid phase stress}

\abbreviations{IR,NMR,UV}

\begin{document}







\begin{abstract}
Gas-particle flows are commonly simulated through two-fluid model at industrial-scale. However, these simulations need very fine grid to have accurate flow predictions, which is prohibitively demanding in terms of computational resources. To circumvent this problem, the filtered two-fluid model has been developed, where large-scale flow field is numerically resolved and small-scale fluctuations are accounted for through subgrid-scale modeling. In this study, we have performed fine-grid two-fluid simulations of dilute gas-particle flows in periodic domains and applied explicit filtering to generate datasets. Then, these datasets have been used to develop artificial neural network (ANN) models for closures such as the filtered drag force and solid phase stress for the filtered two-fluid model. The set of input variables for the subgrid drag force ANN model that has been found previously to work well for dense flow regimes is found to work as well for the dilute regime. In addition, we present a Galilean invariant tensor basis neural network (TBNN) model for the filtered solid phase stress which can capture nicely the anisotropic nature of the solid phase stress arising from subgrid-scale velocity fluctuations. Finally, the predictions provided by this new TBNN model are compared with those obtained from a simple eddy-viscosity ANN model.
\end{abstract}

\section{Introduction}
Gas-particle flows arise in technological applications, e.g., fluidized and circulating fluidized bed reactors, and in nature, e.g., dust storms. There is much interest in studying the characteristics of these flows via mathematical modeling with complementary computer-aided simulations. Reliable modeling and simulations can aid in the design, retrofit, and troubleshooting of industrial processes. As industrial-scale fluidized beds contain trillions of particles, it is impractical to analyze the flow behavior by following the motion of individual particles. In contrast, two-fluid models (TFMs)\cite{Anderson1967,Gidaspow1994}, which treat the fluid and particle phases as inter-penetrating continua, are more viable to analyze and simulate the flows in industrial-scale applications. They have been useful in the analysis of the onset of instabilities in fluidization, and the emergence of inhomogeneous flow structures (e.g., see \cite{glasser1996one,glasser1997fully,glasser1998bubbles}). The TFMs can readily be solved numerically using commercial codes (e.g., ANSYS Fluent), and open-source simulation platforms (e.g., MFIX, OpenFOAM). 

It is now well known that fluidized and circulating fluidized beds manifest structures that span a wide range of length and time scales. The scale of the spatial structures can range from a few particle diameters to the size of the vessel, which can be as large as $10^3 - 10^5$ particle diameters. The macroscale flow structures, referred to as coherent flow structures, can have a large effect on the overall flow hydrodynamics in the device. At the same time, meso-scale structures (such as streamers, clusters and small bubble-like voids) are also important as they affect the emergence of the macroscale structures. As a result, accurate simulations of TFM equations often require fine spatial resolution down to the scale of a few particle diameters\cite{Agrawal2001,Parmentier2011,Ozel2013,sundaresan2018toward} which, in turn, require very small time steps as well. Such highly resolved simulations of industrial-scale processes are not feasible due to the high computational demands\cite{sundaresan2000modeling}. This consideration led to the development of filtered two-fluid models (fTFMs)\cite{Igci2008,Parmentier2011,Ozel2013,Schneiderbauer2017,Sundaresan2018} where the TFM model equations are filtered by a convolution kernel as in the development of Large Eddy Simulation (LES) equations for turbulent flows by averaging the Navier-Stokes equations\cite{pope2001turbulent}. 

The fTFM equations contain several terms representing the consequences of subfilter-scale fluctuations on the spatiotemporal evolution of the filtered variables. Similar to the LES modeling, the importance of subgrid terms in the fTFM equations could be studied by \textit{a priori} tests. It has been shown through the budget analysis of the filtered solid momentum balance generated by filtering fine-grid simulations\cite{Parmentier2011,Ozel2013,Schneiderbauer2017} that the correction to the fluid-particle drag force is of principal importance for gas-particle flows with high mass loading of particles because of the subfilter-scale inhomogeneous distribution of particles. The solid phase stress associated with the subfilter-scale particle velocity fluctuations is of secondary importance \cite{Ozel2013}, while all the other corrections are essentially negligible. As a result, the literature on fTFM model development has focused primarily on the correction to the fluid-particle drag force and, to a lesser extent, on the (filtered) solid phase stress. 

The drag correction models accounting for the effects of unresolved drag due to particle clustering at meso-scale in the literature can be classified as follows. The Energy Minimization Multi Scale (EMMS) model\cite{Li1994,Lu2009,pakseresht2023critical} describes subgrid structures through a heterogeneous index, which is used to estimate the effective drag force. In the framework of fTFM, the explicit correlations were proposed by~\citet{Igci2008,Milioli2012a,Sarkar2016} for the filtered drag in terms of the filtered variables and the filter size. \citet{Parmentier2011} argued that in the presence of subfilter-scale (aka subgrid-scale, as the filter size is usually the same as the grid size in coarse simulations of the fTFM equations) inhomogeneities, the average gas velocity seen by the particles is not the same as the filtered gas velocity and expressed the drag force correction in terms of a subgrid quantity known as the drift velocity. Algebraic models for the drift velocity have been proposed in several studies\cite{Parmentier2011,Ozel2013,Ozel2016}. \citet{Rauchenzauner2022} expressed the drift velocity in terms of the subgrid turbulent kinetic energy of the gas and the scalar variance of the particle volume fraction, which were determined by solving corresponding dynamic transport equations. In a recent study, \citet{Hardy2023} found that the drift velocity could be expressed in terms of the scalar variance of the particle volume fraction, with the same model applying to all filter sizes. 

Several research groups have applied machine learning (ML) techniques to arrive at models for the filtered drag force. 
\citet{Jiang2018} developed the transport equation for the drift velocity and performed a budget analysis of the terms in the developed equation to analyze their importance. They concluded that an algebraic model for the filtered drift velocity would be sufficient for dense fluidized beds. The algebraic model relates the filtered drift velocity to the filtered gas phase pressure gradient and solid volume fraction, and the difference between the filtered gas and solid phase velocities (referred to as the slip velocity). These variables are taken as physics-inspired inputs to an artificial neural network (ANN) model (specifically, a multi-layer perceptron, (MLP)) for the drift flux (which is the product of filtered drift velocity and the filtered solid volume fraction), for given filter size and physical properties of the gas and particles. The drift flux was then used to predict the filtered drag force, as illustrated by~\citet{Parmentier2011}. \citet{Jiang2021} extended the analysis to include the filter sizes and the Reynolds number based on the terminal settling velocity of the particle as additional inputs so that the ANN model can be used for a wide range of fluidized bed applications. Interestingly, \citet{Jiang2018} concluded that a good correlation could be obtained with their ANN model only when the output variable was chosen to be the drift flux; in contrast, their ANN model performed poorly when the drag correction or drift velocity was used as the output variable.  In a very similar context, \citet{Zhang2020} found that an ANN model where the hidden layers included a combination of convolutional layers and fully connected layers revealed better predictions for \textit{a priori} tests of the filtered drag. All these studies considered systems without inter-particle forces; \citet{Tausendcshoen2023} report that the strength of the drag correction is affected by inter-particle forces and proposed an ANN model including the Bond number as a measure of the attractive inter-particle forces. 
In the Eulerian-Lagrangian approach, \citet{Lu2022} developed a filtered drag force ANN model from fine-grid CFD-DEM simulations of dense fluidized beds and they coupled this new drag correction model with the MFiX software for large-scale simulations.
For a more extensive review of ML-based modeling efforts in multiphase flow reactors, the reader is referred to the recent study by \citet{Zhu2022}.

In spite of the progress in the application of ML methods for formulating subgrid drag models for gas-solid flows without the inter-particle forces, unanswered questions remain. The first objective of the present study is to address the following questions:
\begin{itemize}
    \item Why have prior studies found better predictions with some output variables (namely, the drift flux) but not others? 
    \item Is this observation related to the underlying physics of the problem or the use of sub-optimal neural networks? 
    \item Will the input variables for the ANN model identified by \citet{Jiang2021} using datasets generated through simulations of dense fluidized beds be sufficient to model the filtered drift flux in dilute flows such as those encountered in risers?
\end{itemize}  
As noted above, the filtered solid phase stress associated with the subfilter particle velocity variations is the second most important correction. Although the budget analysis found this stress term to be secondary, it could play a role in capturing correctly the smaller-scale structures resolved in the fTFM simulations, which in turn could affect the emergence of the macroscale structures in some flow problems. \citet{Milioli2012a,Ozel2013,schneiderbauer2018validation} proposed an isotropic model for the stress which resembled the Smagorinsky model for the stress in single-phase turbulent flows, and advanced explicit functional models for the pressure and effective viscosity associated with subfilter-scale fluctuations in the velocities of both phases. The importance of anisotropy has been discussed by several researchers\cite{Sarkar2016,Cloete2018a,Cloete2018b,Rauchenzauner2020}. \citet{Rauchenzauner2020} have proposed a dynamic multiphase turbulence model for coarse-grid simulations, which includes transport equations for the scalar variance of the solid volume fraction and the individual components of the turbulent kinetic energies of both phases, requiring additional closure models. These extra transport equations improve predictions but add to the computational cost. Thus, as a second objective of the present study, we explore the use of tensor-based neural network models, which have found use in single-phase turbulent flows\cite{ling2016reynolds}, to constitute the solid phase stress. The predictions offered by this new Galilean invariant model are also compared with simple eddy-viscosity ANN models, similar to earlier proposals in the literature \cite{Ouyang2021}.
Finally, a sample dataset and Python ML model source codes are placed in the GitHub repository (\url{https://github.com/bahardy/fTFM\_ANN\_modeling.git}) for broader use.

The rest of the paper is organized as follows. First, the filtered two-fluid model equations and the subgrid terms to be modeled are briefly recalled. Then, the flow configuration and the filtering procedure used to generate datasets are detailed. Subsequently, the Artifical Neural Network architecture chosen to predict the filtered drag force is presented and the results obtained by this model are discussed. Finally, we introduce the Tensor Basis Neural Network architecture for the particle phase stress, we present the predictions yielded by this more advanced model and compare them with a simple eddy-viscosity approach. The paper ends with a summary of the achievements of this work and suggestions for future research. 

\subsection{Filtered Two-Fluid Model}

As detailed in the work of Igci et al. \cite{Igci2008} and others\cite{Zhang2002, Parmentier2011, Ozel2013}, the filtering of the  mass and momentum balance  equations of the ``micro-scale'' TFM leads to the following set of equations for the gas and solid phases:
\begin{align}\label{eq:fTFMEqs}
\frac{\partial}{\partial t}\left(\rho_g \bar \phi_g \right) + \nabla \cdot (\rho_g \bar \phi_g \tilde \u_g) & = 0 \\
\frac{\partial}{\partial t}\left(\rho_s \bar \phi_s \right) + \nabla \cdot (\rho_s \bar \phi_s \tilde \u_s) & = 0 \\
\frac{\partial}{\partial t}\left(\rho_g \bar \phi_g \tilde \u_g \right) + \nabla \cdot (\rho_g \bar \phi_g \tilde \u_g \tilde \u_g) & = - \nabla \cdot \boldsymbol{\sigma}_g - \overline {\phi_g \nabla p_g} - \nabla \cdot \overline{\boldsymbol{\Sigma}}_g^d - \overline{\mathbf I}_{gs} + \rho_g \bar \phi_g \mathbf g \\
\frac{\partial}{\partial t}\left(\rho_s \bar \phi_s \tilde \u_s \right) + \nabla \cdot (\rho_s \bar \phi_s \tilde \u_s \tilde \u_s) & = - \nabla \cdot \boldsymbol{\sigma}_s - \overline {\phi_s \nabla p_g} - \nabla \cdot \overline{\boldsymbol{\Sigma}}_s + \overline{\mathbf I}_{gs} + \rho_s \bar \phi_s \mathbf g 
\end{align}

Here, $\rho_g$, $\rho_s$ are the gas and solid phase densities, respectively, $p_g$ is the gas phase pressure, and $\mathbf g$ is the gravitational acceleration. The filtered volume fractions for gas ($k=g$) and solid ($k=s$) phases are defined as
\be
\bar \phi_k(\bx, t) = \int_V \phi_k(\by, t)G(\by-\bx) d\by, \quad k=g,s
\ee
where $\phi_k(\bx, t)$ is the volume fraction for each phase given by the ``micro-scale'' TFM, $G(\br)$ is the filter convolution  kernel satisfying $\int_V G(\br) d\br = 1$.
Similar to LES of compressible flows \cite{Garnier2009}, the phase velocities are filtered through the Favre-averaging as
\be
\tilde \u_k(\bx, t) = \frac{1}{\bar \phi_k}\int_V \phi_k(\by, t) \u_k(\by,t) G(\by-\bx) d\by = \frac{\overline{\phi_k \bu_k}}{\bar \phi_k}, \quad k=g,s.
\ee
The filtered quantity for each phase, denoted with a bar, in Eq. \eqref{eq:fTFMEqs} is defined as:
\be
\bar{\mathcal{Q}_k}(\bx, t) = \int_V \mathcal{Q}_k(\by, t)G(\by-\bx) d\by, \quad k=g,s
\ee
where $\mathcal{Q}_k(\by, t)$ is a quantity for each phase in the ``micro-scale'' TFM. The filtered and its fluctuating quantities are described as 
\be
\mathcal{Q}_k(\bx, t) = \bar{\mathcal{Q}_k}(\bx, t) + \mathcal{Q}_{k}^{'}(\bx, t). 
\ee
$\bar{\Sigma}_g^d$ and $\bar{\Sigma}_s$ are respectively the deviatoric filtered gas phase and total solid phase micro-scale stresses. $\bar{\mathbf I}_{gs}$ is the  filtered gas-solid interphase momentum exchange term; in gas-particle systems, it is principally only the drag force. The explicit expressions for the filtered micro-scale stress tensors can be found in earlier works\cite{Igci2008,  Parmentier2011,Ozel2013}. 

The filtered drag force term is defined as 
\be
\bar {\mathbf I}_{gs} = \overline{\beta\left(\bu_g - \bu_s \right)},
\ee
where $\beta$ is the microscopic drag coefficient (also defined as $\beta = \mathlarger{\frac{\rho_s \phi_s}{\tau_p}}$, with $\tau_p$ the particle relaxation time \cite{Parmentier2011,Ozel2013}).
In the literature, it is very common practice \cite{Li1994, Zhang2002, Lu2009, Wang2010, Schneiderbauer2013e, Schneiderbauer2015b, Schneiderbauer2017, Tausendcshoen2023} to account for the drag correction required in the fTFM by introducing an effective drag coefficient $\beta_e$, namely,    
\be
\bar {\mathbf I}_{gs} \simeq \beta_e \left(\tilde \bu_g - \tilde \bu_s \right),
\label{eq:effective_drag_coefficient}
\ee
where $\beta_e$ has to be determined for fTFM simulations. The relation between the effective drag coefficient and the microscopic drag coefficient computed from the filtered quantities, here noted $\tilde \beta$, is usually expressed in terms of a drag correction factor $H_d$ \cite{Lu2009} defined as
\be
H_d =  \frac{\beta_e}{\tilde \beta}.
\label{eq:correction_factor}
\ee

Numerous studies have sought to improve the functional description of $H_d$ \cite{Igci2008, Igci2011,Parmentier2011, Ozel2013, Milioli2012a, Sarkar2016}. It must be stressed that Eqs. \eqref{eq:effective_drag_coefficient} and \eqref{eq:correction_factor} implicitly assume that the required drag correction (i.e. the subgrid drag force term) is aligned with the filtered slip velocity and that the drag correction factor is isotropic. (Note: the terms subgrid and subfilter are used interchangeably as in fTFM simulations the filter size is usually taken to be the grid size.) 
As noted by \citet{Tausendcshoen2023}, a more general model would be 
\be
\bar {\mathbf I}_{gs} = \mathbf H_d \tilde \beta \left(\tilde \bu_g - \tilde \bu_s \right), 
\label{eq:correction_factor_tilde}
\ee
where $\mathbf H_d$ is a second-order tensor. 
Nevertheless, it has been verified \cite{Parmentier2011,Ozel2013, Schneiderbauer2018} that Eq. \eqref{eq:correction_factor} is adequate and that, to a very good approximation, the filtered drag force can be written as 
\be
\bar {\mathbf I}_{gs} = \tilde \beta \left(\tilde \bu_g + \bv_d - \tilde \bu_s \right),
\label{eq:filtered_drag}
\ee
where $\bv_d$ is the so-called drift velocity, defined as 
\be
\bar \phi_s \bv_d = \overline{\phi_s \bu_g} - \bar \phi_s \tilde \bu_g. 
\label{eq:drift_flux}
\ee
In this paper, the product $\bar \phi_s \bv_d$ will be referred to as the ``drift flux''. The filtered drag and the drift flux can be explicitly linked as follows
\be
\bar {\mathbf I}_{gs} = \frac{\rho_s \bar \phi_s }{\tilde \tau_p} \left(\tilde \bu_g - \tilde \bu_s \right) + \frac{\rho_s  }{\tilde \tau_p} \bar \phi_s \bv_d.
\label{eq:filtered_drag_drift_flux}
\ee
where $\tilde \tau_p$ is the particle relaxation time computed from filtered quantities. 
Most functional models proposed in the literature considered $\bv_d$ to be aligned with the resolved slip velocity \cite{Parmentier2011, Ozel2013} (which eventually boils down to an expression of the form given by Eq. \eqref{eq:effective_drag_coefficient}) though, Ozel et al. \cite{Ozel2017} suggested that the drift velocity has an additional, albeit small, explicit dependence on the subgrid variance of the solid volume fraction. The subgrid variance of the solid volume fraction also enters the dynamic drift velocity model of  \citet{Rauchenzauner2019} while \citet{Hardy2023} recently proposed an explicit model to deduce the drift velocity from the subgrid variance, independently of the filter size.

The second most important sub-grid contribution in the fTFM comes from the filtered solid phase (or meso-scale) stresses, defined as
\be
\boldsymbol \sigma_s= \overline{\rho_s \phi_s \u_s \u_s} - \rho_s \bar \phi_s \tilde \u_s \tilde \u_s.
\label{eq:meso-scale_stresses}
\ee
While the drag correction has been found to saturate as the filter size increases, the meso-scale stress term grows monotonically with the filter size \cite{Ozel2013}. Its modeling is therefore becoming increasingly important as the mesh resolution is lowered for very large scale simulations. Prior studies \cite{Agrawal2001, Andrews2005, Igci2008} have concluded that the filtered solid phase micro-scale stresses ($\overline{\boldsymbol \Sigma}_s$) described by the kinetic theory of granular flows is much weaker than the meso-scale stresses, even for moderately large filter sizes (typically filter sizes larger than 15-20 particle diameters).

Finally, the filtered pressure gradient term can be decomposed as the sum of a resolved and a subgrid term as 
\be
\overline {\phi_k \nabla p_g} = \bar \phi_k \nabla \bar p_g + \Phi_k^{sgs},  \hspace{1cm} k=g,s, 
\ee
Some authors~\cite{DeWilde2005, Zhang2002} proposed to model $\Phi_k^{sgs}$ as an added mass term, but most studies concluded that this term was small with respect to the filtered drag and meso-scale stress terms and that it is sufficient to retain only the resolved part of the pressure gradient, i.e. $\overline {\phi_k \nabla p_g} \simeq \bar \phi_k \nabla \bar p_g$.

\subsection{Flow Configurations and Generating Dataset through TFM simulations}

A prerequisite for the training and \textit{a priori} validation of ANN models for the filtered drag and filtered stresses is the generation of a database of fine-grid TFM simulation results covering a wide range of physical parameters. As noted earlier, \citet{Jiang2021} performed fine-grid simulations of dense fluidized beds.  Their datasets had only a sparse representation of dilute flow conditions. Therefore, their findings apply to  flow conditions with particle volume fractions exceeding ~0.1 and become less accurate at more dilute conditions. One of the goals of the present study is to examine if the input variables required for the ANN model for drag correction are any different for dilute flow conditions. For that reason, we have performed a number of dilute flow simulations using the two(multi)-fluid solver \textit{neptune\_cfd}~\cite{Neau2020}. 

The computational domain is tri-periodic with the gravitational acceleration acting along the $z-$direction. A source term mimicking an external pressure gradient is added to the gas and solid phase momentum transport equations to compensate for the weight of the mixture against the gravitational acceleration. The different investigated cases and their physical parameters (particle diameter $d_p$, density $\rho_s$ and domain-averaged solid fraction $\left< \phi_s\right>$) are summarized in Table~\ref{tab:parameters}. The following variables are fixed through all simulations: the gas density $\rho_g = \SI{1.2}{\kg\per\cubic\m}$, the gas dynamic viscosity  $\mu_g = \SI{1.8e-5}{\pascal\s}$ and the particle restitution coefficient $e_c = 0.9$. We also report the particle Froude number $\Fr_p = \mathlarger{\frac{U_t^2}{g d_p}}$, the particle Reynolds number based on terminal settling velocity $\Rey_p = \mathlarger{\frac{\rho_g U_t d_p}{\mu_g}}$ and the Froude number based on the grid size $ \Fr_{\bar \Delta} = \mathlarger{\frac{g\Delta}{U_t^2}}$ where $U_t$ is the particle terminal settling velocity estimated from Wen \& Yu drag law \cite{Wen1966} in Table~\ref{tab:parameters}.
The mesh size $\Delta$ is uniform, with 640 cells along the vertical direction, and the aspect ratio between the vertical and horizontal dimensions of the domain is 4. 

\begin{figure}[h!]
    \centering
    \includegraphics[width=0.5 \textwidth]{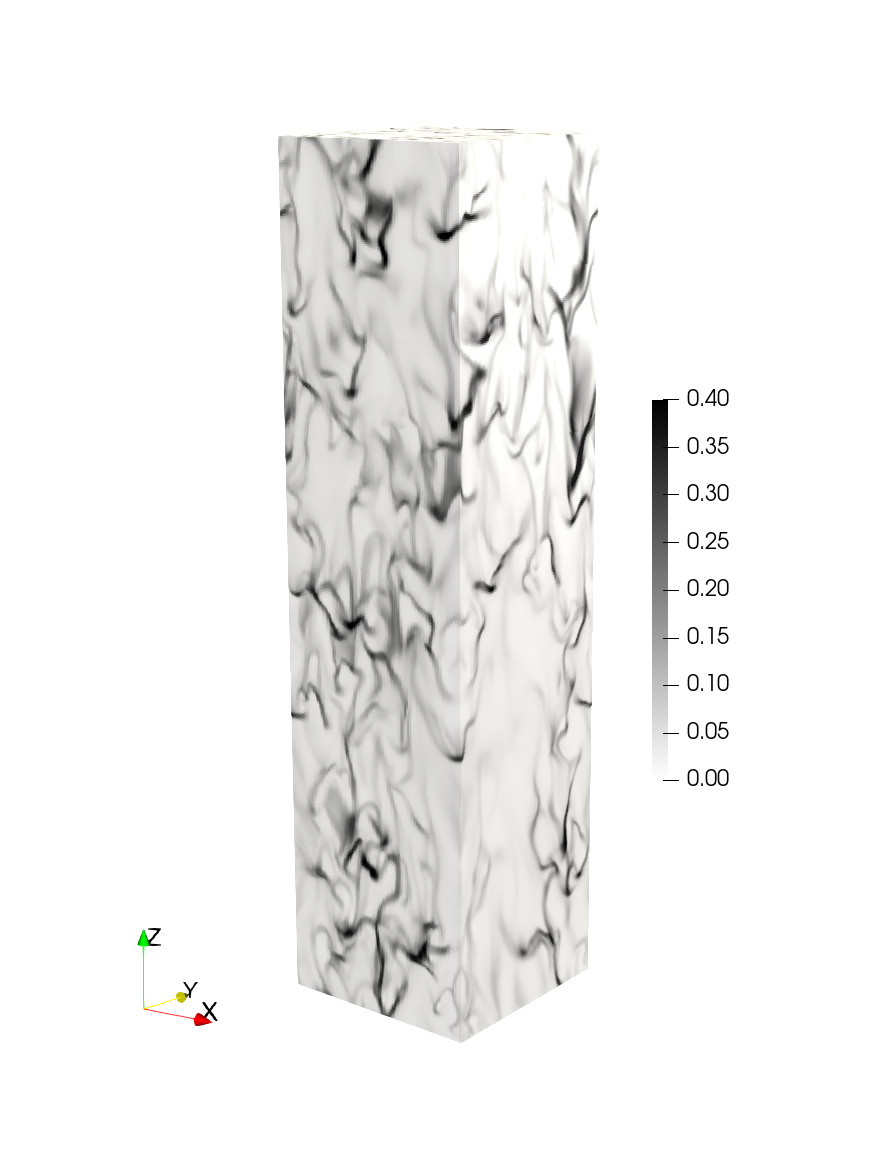}
    \caption{Instantaneous solid volume fraction $\phi_s$ in a TFM simulation of a tri-periodic fluidized bed in Case 1 (see Table~\ref{tab:parameters}). }
    \label{fig:phi_snapshot}
\end{figure}

In the remainder of this study, Case 1 will be considered as the reference case as it corresponds to typical conditions for the fluidization of Geldart A particles with air. It has to be emphasized that this case has already been studied extensively in the literature. A snapshot of the solid volume fraction field from Case 1 is shown in Figure \ref{fig:phi_snapshot}, highlighting the formation of typical elongated clusters. 
\begin{table}[h!]
\centering
\begin{tabularx}{\textwidth}{p{2cm}|p{2cm}p{2cm}p{2cm}p{2cm}p{2cm}p{2cm}}
\hline
&  $d_p \, (\si{\micro\m})$ & $\rho_s \, (\si{\kg\per\cubic\m})$ & $\left<\phi_s\right>$ &  $\Fr_p$ & $\Rey_p$ & $\Fr_{\Delta}$ \\[6pt]
\hline
Case 1 & 75 & 1500 & 0.05 & 65.33 & 1.10& 20.41 \\[5pt]
Case 2 & 90 & 1500 & 0.05 & 101.09& 1.79& 20.41 \\[5pt]
Case 3 & 100 & 1500 & 0.05 & 128.37 & 2.37 & 20.41 \\[5pt]
Case 4 & 75 & 3000 & 0.05 & 228.50 & 2.05 & 20.41 \\[5pt]
Case 5 & 75 & 1500 & 0.10 & 65.33 & 1.10 & 20.41 \\[5pt]
Case 6 & 150 & 2500 & 0.05 & 663.0 & 9.88 & 7.60 \\[5pt]
Case 7 &  75 & 1500 & 0.08 & 65.33 & 1.10  & 20.41 \\[5pt]
Case 8 & 150 & 2500 & 0.10 & 663.0 & 9.88 & 7.60 \\[5pt]
Case 9 & 120 & 2000 & 0.05 & 309.96 & 4.83 & 45.80 \\[5pt]
\hline
\end{tabularx}
\caption{Physical and flow parameters of the fine-grid TFM simulations of a gas-solid unbounded fluidized bed in a tri-periodic domain.}
\label{tab:parameters}
\end{table}  

\subsection{Filtering of fine-grid TFM data}
Filtered and subgrid terms in the fTFM equations can be computed by applying an explicit filter on the fine-grid simulation data~\cite{Igci2008}. In the present work, we use a box (or top-hat) filter $\bar G$ in the physical space: 
\be
\bar G(\mathbf r) = \left\lbrace 
\begin{array}{ll}
    \mathlarger{\frac{1}{\bar \Delta^3}},  & |r_i| < 0.5 \bar \Delta , \; i=x,y,z \\
    0, & \text{otherwise},
\end{array}
\right.
\label{eq:tophat_filter}
\ee 
where $\bar \Delta$ is the filter width. By doing so, the filtered drag, the drift velocity (and drift flux), and the filtered solid phase stresses have been computed from Eqs. \eqref{eq:filtered_drag}, \eqref{eq:drift_flux} and \eqref{eq:meso-scale_stresses}, respectively, for a range of filter-to-grid size ratios: $\mathlarger{\frac{\bar \Delta}{\Delta}} = [2,4,6,8,10,12,16]$. 

To address the first objective, we build new ANN-based models for the filtered drag force and solid stresses from a set of well-chosen markers. The selection of those markers should be inspired by underlying physics and by the many studies that have sought explicit coarser-grained models for fTFM. The case of the filtered drag force and solid stresses will be addressed separately later in more details, but a few key quantities can already be identified. The filtered slip velocity 
\be
\tilde \bu_{slip} = \tilde \bu_g - \tilde \bu_s,
\ee
appears in numerous functional models next to the filtered solid volume fraction $\bar \phi_s$ and the filter size $\bar \Delta$ (see \citet{Cloete2018c} for a comparative study of different existing models). It has been more recently observed by \citet{Jiang2018} and confirmed by later studies \cite{Jiang2021, Zhang2020, Zhu2020, Ouyang2023} that the addition of the filtered pressure gradient acting against the gravitational acceleration  $\mathlarger{\frac{\partial \bar p_g}{\partial z}}$ as an input to their filtered drag ANN model dramatically improved their results. Based on that, \citet{Jiang2020} formulated an explicit drift velocity model with an additional dependence on the filtered pressure gradient. Besides, recent studies~\cite{Ozel2013,Ozel2017,Rauchenzauner2019,Hardy2023} identified the subgrid variance of the solid volume fraction as a potentially good marker for the filtered drag. The underlying idea is that the drift velocity (and therefore the drag correction) originates from a heterogeneous distribution of the particles at the subgrid scale and that the subgrid variance of the particle volume fraction (henceforth, simply referred to as SV) is a good measure of this level of heterogeneity. In this study, the SV will be defined as 
\be
\overline{\phi'^2} = \overline{\phi^2} - \overline{\phi}^2.  
\ee
The SV was previously introduced for filtered drag force modeling by Schneiderbauer \cite{Schneiderbauer2017} where it appears in the expression of their drag correction factor. 
Their model additionally depends on the subgrid correlated kinetic energy of the solid phase, $k_s^{sgs}$, defined as 
\be
k_s^{sgs} = \frac{1}{2}\mathrm{tr}(\boldsymbol \sigma_{s}). 
\ee

As for the filtered solid stresses modeling, single phase turbulence models \cite{smagorinsky1963, yoshizawa1986} and previous functional modeling efforts for gas-solid flows \cite{Ozel2013, Milioli2012a, schneiderbauer2018validation} suggest that the deviatoric part of the filtered rate-of-deformation tensor, defined as
\be
\tilde {\mathbf S}_s = \frac{1}{2}(\nabla \tilde \u_s + \nabla \tilde \u_s^T) - \frac{1}{3} \nabla\cdot \tilde \u_s \mathbf I 
\label{eq:strain_rate}
\ee
should definitely play a role. The rotation-rate tensor 
\be
\tilde {\mathbf R}_s = \frac{1}{2}(\nabla \tilde \u_s - \nabla \tilde \u_s^T) 
\label{eq:rotation_rate}
\ee 
will also be involved in the ANN modeling of the filtered stresses later on.
Overall, the gradient of the phase-filtered velocities and of the filtered solid volume fraction $\nabla \bar \phi_s$ have been extracted from our fine-grid TFM simulation results as they contain non-local information that can potentially improve the description of the filtered drag and filtered solid stresses.

\section{Neural Network Modeling of Filtered Drag Force}

The first study to exploit the neural network approach for modeling the filtered drag (FD) force was proposed by \citet{Jiang2018}. These authors first developed the transport equation of the drift velocity. They used the fine-grid TFM simulation results of a bubbling fluidized bed with Geldart-A type particles to assess the importance of unclosed terms in the transport equation. The transport equation was then simplified to the algebraic model, which was closed by using a 3-marker model using the filtered solid volume fraction $\bar \phi_s$, the filtered slip velocity $\tilde u_{slip,z}$ and the gas phase gradient $\mathlarger{\frac{\partial \bar p_g}{\partial z}}$ in the gravitational acceleration direction. Instead of proposing an explicit functional form for the 3-marker model, they used an MLP made of 3 hidden layers of 128, 64 and 32 nodes to close the component of the drift flux aligned with gravitational acceleration, i.e. $\bar \phi_s v_{d,z}$. The filtered drag force was then deduced from an explicit relation derived from Eq. \eqref{eq:filtered_drag}. These authors observed that the inclusion of the filtered gas phase pressure gradient significantly improved the degree of correlation of their model with the exact filtered drag. Different neural networks were trained for each filter size, so that the filter size was not considered as a distinct marker at that point. They were unable to achieve a similar level of correlation when the filtered drag force or the drift velocity was set as output variable of their ANN, instead of the drift flux.
In a later study \cite{Jiang2021}, the same group verified that a filter-size dependent ANN model could be applied to large-scale simulations. In addition, they extended the range of application of their filtered drag force model by training their ANN on a wide range of physical parameters and by using appropriate scaling for the input markers. They also concluded that the particle Reynolds number or the Archimedes number should come as an additional marker to account for variations in physical properties. 
The prediction improvement offered by the addition of the filtered gas phase pressure gradient was also confirmed by \citet{Ouyang2023} through interpretable ML metrics.
\citet{Zhang2020} proposed a convolutional neural netwok (CNN) architecture to predict the filtered drag in periodic unbounded gas-solid flows. Information of neighboring grid points was therefore inherently included in their filtered drag force model by the structure of the network and the authors report increased performance with respect to MLP models or explicit dynamic models. They also concluded that the filtered gas phase pressure gradient improves the predictions of their model but to a lesser extent that with the MLP architecture since information from neighboring cells is already embedded in the model. It is however expected that such CNN model will be more computationally intensive and therefore less tractable for practical fTFM simulations at industrial-scale. \citet{Zhu2020} compared a classical MLP and a gradient boosting framework to predict the filtered drag and integrated these ML models into fTFM simulations of bubbling and turbulent fluidized beds. They subsequently validated their ANN-assisted fTFM simulations against experimental results of a bubbling fluidized bed and found satisfactory agreement between the two approaches. 
Very recently, \citet{Tausendcshoen2023} used Eulerian-Lagrangian simulation results to propose distinct ANN models for the different components of the drift flux. These authors found that these anisotropic drag models lead to better results that the isotropic counterpart, as previously noticed by \citet{Cloete2018a, Cloete2019} for explicit models. These authors also added the Bond number as an additional marker to account for cohesive effects in gas-solid flows.


\subsection{Artifical Neural Network (ANN) Architecture for Filtered Drag Force} 
\label{section:MLP_architecture}
\begin{figure}
    \centering
    \tikzset{every picture/.style={scale=0.7}}
    \input{MLP_filteredDrag_drawing}
    \caption{Multi-layer perceptron (MLP) architecture for filtered drag force modeling.}
    \label{fig:FD_NN_arch}
\end{figure}
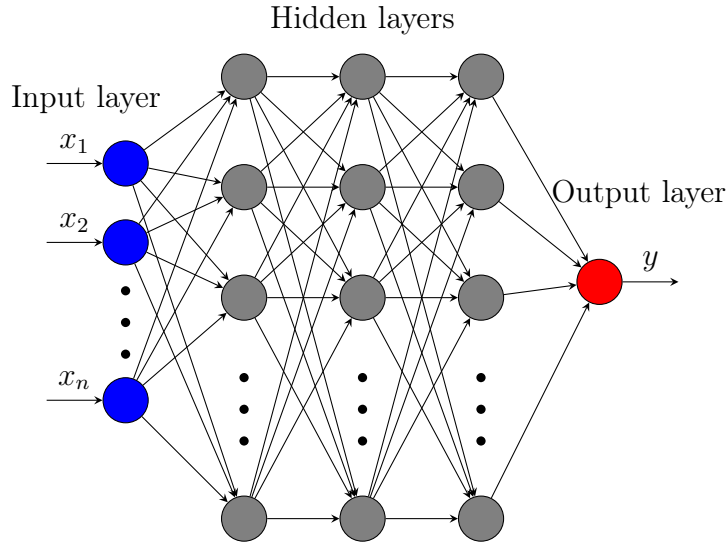

To address whether the many different markers used in the previous NN models represent the underlying physics, or  the use of sub-optimal neural networks, 
we developed our NN architecture, which is a feedforward MLP similar to \citet{Jiang2018}. 
This neural network architecture is sketched in Figure \ref{fig:FD_NN_arch}. It consists of one input layer, some hidden layers and one output layer. The input layer corresponds to the physical markers selected to predict the target quantity. Hidden layers are made of a number of nodes or \textit{neurons}. Because the network is densely connected, each node takes as input all the nodes in the previous layer. The outcome of a single neuron $i$ within a layer $n$ is described by 
\be
z_i^{(n)} = f(\mathbf w^T \mathbf z^{(n-1)} + w_0) 
\ee
where $\mathbf z^{(n-1)}$ is the output vector of layer $n-1$, $\mathbf w$ is the weight vector of the node, $w_0$ is the bias and $f$ is the activation function. The number of hidden layers and the numbers of nodes per layer are two hyperparameters of the problem that are discussed further below. The output layer is made of a single node whose value (the target quantity $y$) should allow us to compute the filtered drag in a straightforward manner, i.e. using an explicit model. 

Using the datasets generated for dilute systems in the present study, To that end, we investigated the following strategies: 
\begin{enumerate}
\item develop an ANN model for the filtered drag force (FD-ANN): the target quantity $y$ is the filtered drag $\bar {\mathbf I}_{gs}$ 
\item develop an ANN model for the drift velocity (DV-ANN): the target quantity is the drift velocity $\bv_d$ and one uses an explicit expression similar to Eq. \eqref{eq:filtered_drag} to compute the filtered drag force, and
\item develop an ANN model for the drift flux (DF-ANN): the target quantity is the drift flux $\bar \phi_s \bv_d$ and one uses an explicit expression similar to Eq. \eqref{eq:filtered_drag_drift_flux} to compute the filtered drag force.


\end{enumerate}

In what follows, the architecture of the ANN for filtered drag force prediction is kept unchanged in order to compare the capabilities of the different aforementioned strategies for the same level of complexity. The current ANN is made of 3 hidden layers of respectively 128, 32 and 8 nodes, with a ReLU (Rectified Linear Unit) activation function. The output layer has a linear activation function to return the regression result. The loss function used to trained the network is the mean absolute error (MAE), defined as 
\be
\mathrm{MAE} = \frac{1}{N}\sum_i^{N}|\hat y_i - y_i|,
\ee
where $N$ is the number of data points used in a training batch.
The dataset for the reference case is made of \num{1.792e6} entries (accounting for the 7 filter widths). Training and testing phases use subsets corresponding to 80\% and 20\% of the dataset, respectively. Among the training dataset, 20\% of the data are further preserved for validation and prevent overfitting during the learning process. 

The prediction accuracy of the developed ANN models will be assessed using the coefficient of determination $R^2$, defined as 
\be
R^2 = 1 - \frac{\sum_i (y_i - \hat y_i)^2}{\sum_i (y_i - \left<y\right>)^2}
\ee
where $y_i$ is the exact value of the target quantity for the $i$th observation (specifically, the target quantity obtained by filtering the fine-grid TFM simulation results), $\hat y_i$ is the corresponding ANN model output value and $\left<y\right>$ is the mean value of $y$ over the dataset. 
The quality of the model can also be quantified by the Probability Density Function (PDF) of the relative error $e$, defined as 
\be
e_i = \frac{(y_i - \hat y_i)}{y_i}.
\ee

\section{Results from the ANN Model Analysis of the Filtered Drag Force}


We begin this section by comparing the filtered drag force results of Jiang et al.\cite{Jiang2018} and  \citet{Rauchenzauner2019} for dense fluidized beds. As these authors employed different sets of input variables to model the filtered drag force, the comparison described below illustrates the non-uniqueness of the choices of input variables to model the filtered drag force. 
We then turn our attention to ANN models for dilute flows based on the simulations performed in our study, where we consider the quality of predictions afforded by three different combinations of input and output variables in the ANN models. 

\subsection{Dense bubbling fluidized bed: different choices of input variables to estimate the drag correction}

Jiang et al. developed an ANN model for the filtered drag force in  dense, bubbling fluidized beds with average solid volume fractions in the range of  0.25 to 0.4 \cite{Jiang2021}. The authors trained the ANN on about 20 different combinations of particle properties in the Geldart-A and A/B groups. In order to obtain a universally applicable model, the ANN model is based on dimensionless input and output variables. For a specified gas-particle system, the ANN model requires the following four input quantities (and hence their model is referred to as 4-marker model): the filter size, the filtered solid volume fraction and the filtered slip velocity, and the component of the filtered gas phase pressure gradient in the gravitational acceleration direction, all of which are available in a fTFM model simulation. They found that this model can be made applicable to different gas-particle systems by including either the particle Reynolds number based on the terminal velocity $\mathrm{Re_{p}}$ or the Archimedes number $\mathrm{Ar} = \mathlarger{\frac{(\rho_s - \rho_g)\rho_g d_p^3 g}{\mu_g^2}}$ as an additional input variable. They reported a model that employed the Reynolds number as the extra input, which we consider here. Their  drift flux ANN model takes the form: 
\be
\frac{\bar\phi_s v_{d,z}}{\phi_{s,\mathrm{max}} U_t} = f\left(\frac{\bar \phi_s}{\phi_{s,\mathrm{max}}}, \frac{\tilde u_{slip,z}}{U_t}, \frac{\bar \Delta}{L_c}, \frac{1}{\rho_s g}\frac{\partial \bar p_g}{\partial z}, \Rey_p\right).
\label{eq:Jiang-ANN_model}
\ee
where the characteristic length scale $L_c$ is set equal to $d_p \Fr_p^{1/3}$, as suggested by \citet{Radl2014}.



\begin{figure}
    \centering
    \includegraphics[width=0.7\textwidth]{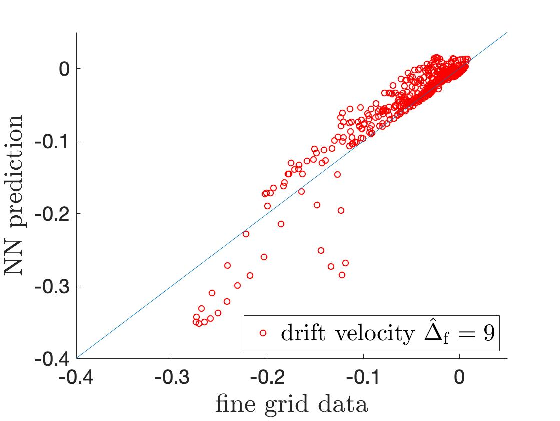}
    \caption{Drift velocity predicted by the ANN of \citet{Jiang2021}, compared to the filtered fine-grid simulation data produced by a different research group~\cite{Rauchenzauner2019} for a dimensionless filter size $\hat{\Delta}_{\rm f} = 9$. The filter-size was made dimensionless using the characteristic length-scale $L_{c} = d_p \Fr_p^{1/3}$. 
    \cite{Radl2014}.}
    \label{fig:comparisonscatter}
\end{figure}

\citet{Rauchenzauner2019} performed fine-grid simulations of a bubbling fluidized bed of Geldart-A type particles, filtered their simulation results and used them to validate a functional model for drift velocity, which used the filtered solid volume fraction, the SV and the  kinetic energy associated with the subgrid  particle phase velocity fluctuations (which they referred to as the turbulent kinetic energy, TKE). As SV and TKE are not available in an fTFM simulation, additional transport equations must be solved to track their spatiotemporal evolution. Nevertheless, it is interesting to see that \citet{Rauchenzauner2019} and \citet{Jiang2021} employed different sets of input variables to model the drift velocity. To test whether the dataset generated by \citet{Rauchenzauner2019} could have been captured by the ANN model supplied by \citet{Jiang2021}, we compared the predictions of the \citet{Jiang2021} model with the data generated by \citet{Rauchenzauner2019}. (Specifically, we tested the results from case 2 of \citet{Rauchenzauner2019} which closely corresponded to case 2 of \citet{Jiang2021}.)  Figure \ref{fig:comparisonscatter} reveals fairly good correlation, suggesting that the drift velocity could be modeled by either combination of input variables. The combination of inputs suggested by \citet{Jiang2021} is perhaps advantageous as it does not require the simulation of additional transport equations. It also suggests that SV and TKE can be estimated in terms of the local quantities employed as input variables by \citet{Jiang2021}; i.e., the transport equations for SV and TKE can be approximated by a local-equilibrium approximation. The filtered slip velocity and the filtered gas phase pressure gradient in the \citet{Jiang2021} model appear in the transport equations for SV and TKE (even when they are simplified with a local-equilibrium approximation).

\subsection{New ANN Filtered Drag Force Models for Dilute-to-Moderately Dense Flows}
\subsubsection{Model for the Reference Case}
The MLP architecture described in Figure \ref{fig:FD_NN_arch} was used to predict the filtered drag force for the Reference Case  (Case 1 in Table \ref{tab:parameters}), involving Geldart A-type particles in the dilute regime. 
We explored different choices of input quantities (markers) to the network to see if conclusions drawn in the dense regime apply to the dilute case as well.
The present analysis focuses on the vertical component of the filtered drag force given its primary importance in fluidization. 

\begin{figure}
    \centering
    \begin{subfigure}{0.45\textwidth}
    \includegraphics[width=\textwidth]{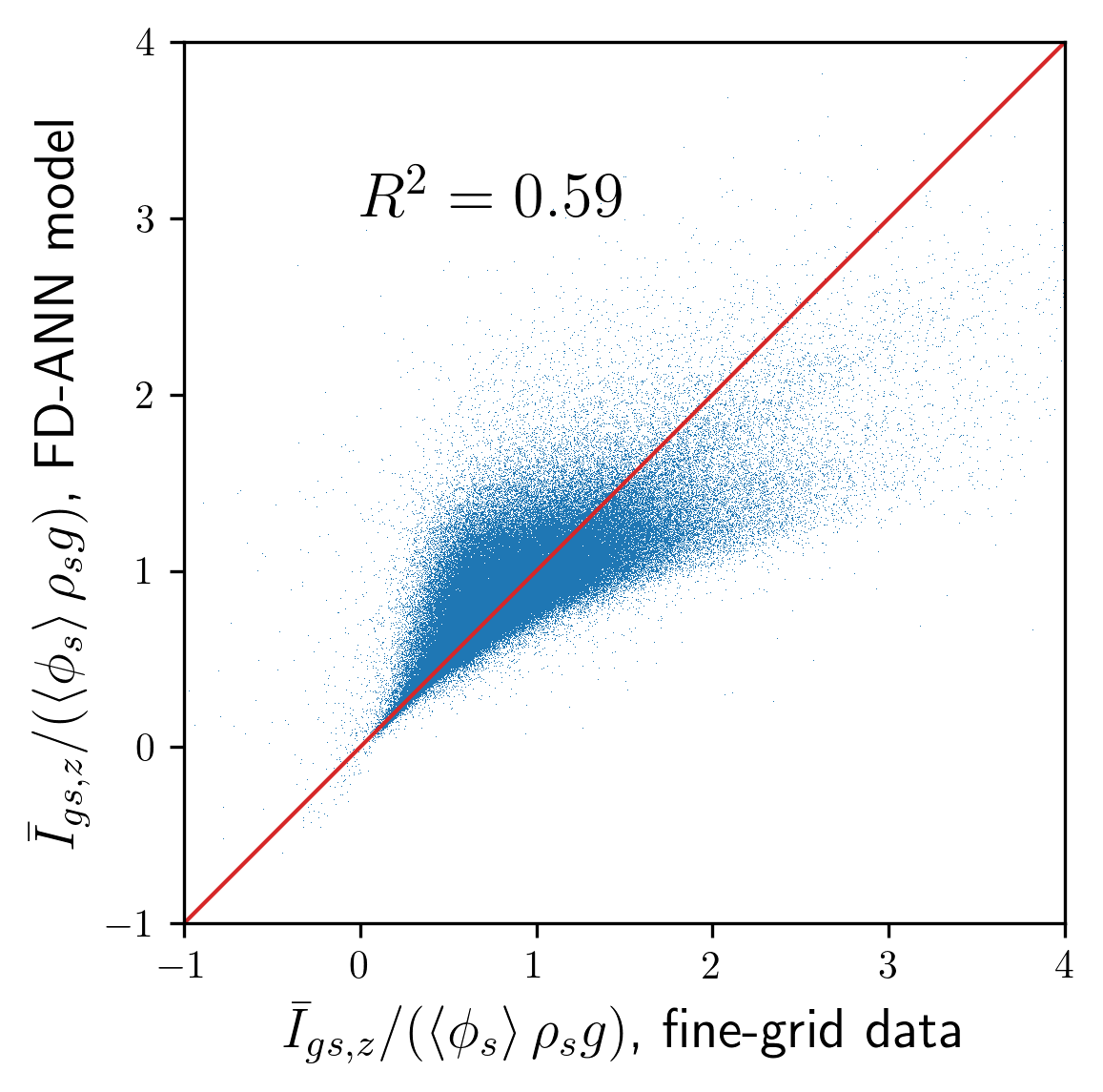}
    \end{subfigure}
    \hfill
    \begin{subfigure}{0.45\textwidth}
    \includegraphics[width=\textwidth]{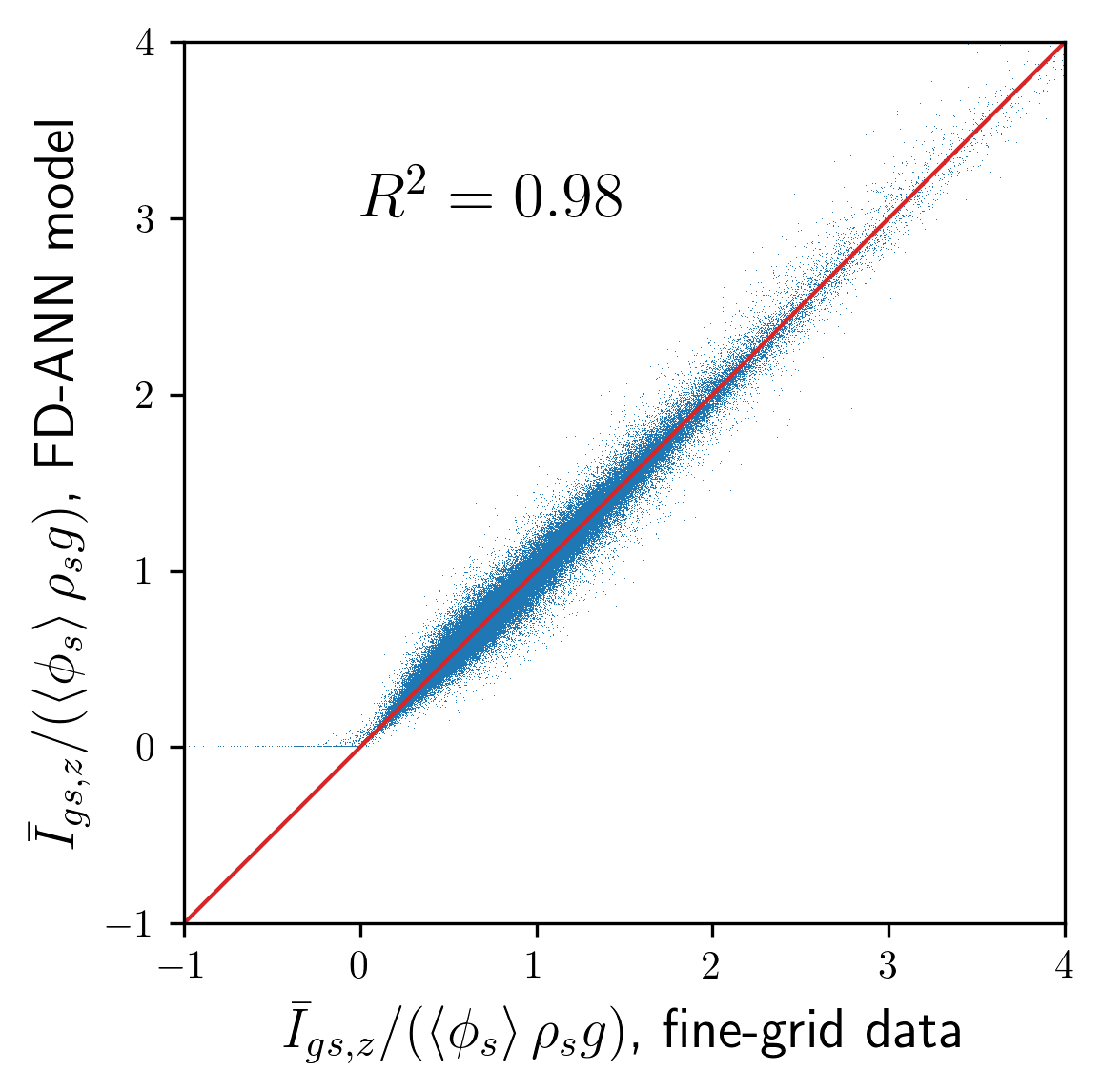}
    \end{subfigure}
    \caption{Assessment of the FD-ANN model predictions. Left: 3-marker model $(\bar \phi_s, \bar \Delta, \tilde u_{slip,z})$. Right: 4-marker model  $(\bar \phi_s, \bar \Delta, \tilde u_{slip,z}, \mathlarger{\frac{\partial \bar p_g}{\partial z}})$.}
    \label{fig:FD-ANN-3vs4marker}
\end{figure}

We start by taking the filtered drag force itself as the output quantity of the network (FD-ANN). Figure \ref{fig:FD-ANN-3vs4marker} compares the predictions of the 3- and 4-marker FD-ANN models, respectively defined by 
\be
\bar I_{gs,z} = f\left(\bar \phi_s, \bar \Delta, \tilde u_{slip,z} \right) 
\ee
and 
\be
\bar I_{gs,z} = f\left(\bar \phi_s, \bar \Delta, \tilde u_{slip,z}, \frac{\partial \bar p_g}{\partial z} \right). 
\ee
In line with \citet{Jiang2018} findings in the dense regime, we observe that the inclusion of the filtered gas phase pressure gradient in the markers of the ANN dramatically improves the accuracy of the model prediction in the dilute flow regime as well. 
\citet{Zhang2020} came to the same conclusion with an MLP architecture, though the authors report lower correlation coefficients, even when the filtered gas phase pressure gradient is added. 
\citet{Jiang2018} found that FD-ANN model performed poorly for dense flows, which is clearly not the case for dilute flows. 

\begin{figure}
    \begin{subfigure}{0.45\textwidth}
    \includegraphics[width=\textwidth]{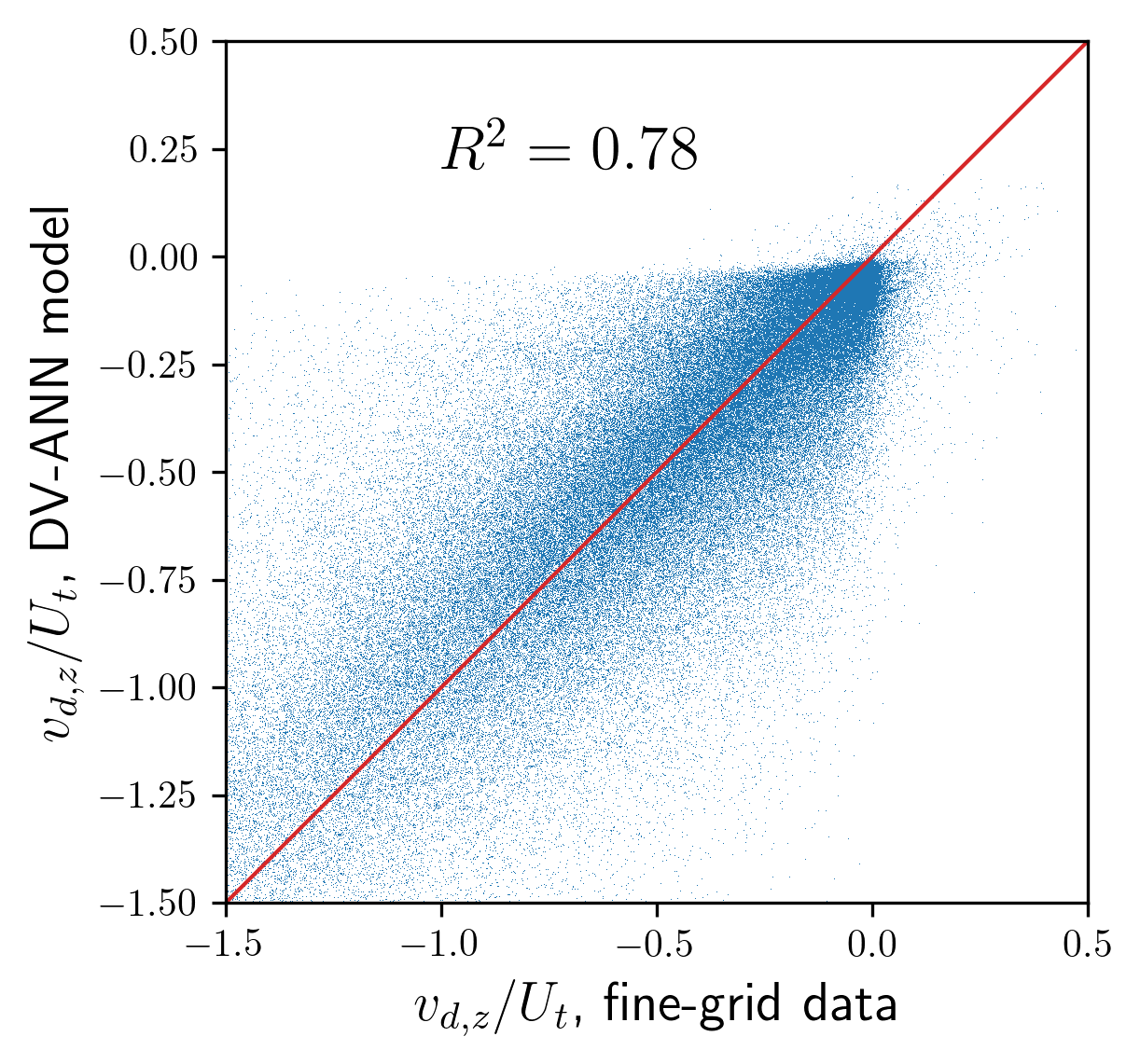}
    \end{subfigure}
    \hfill
    \begin{subfigure}{0.45\textwidth}
    \includegraphics[width=\textwidth]{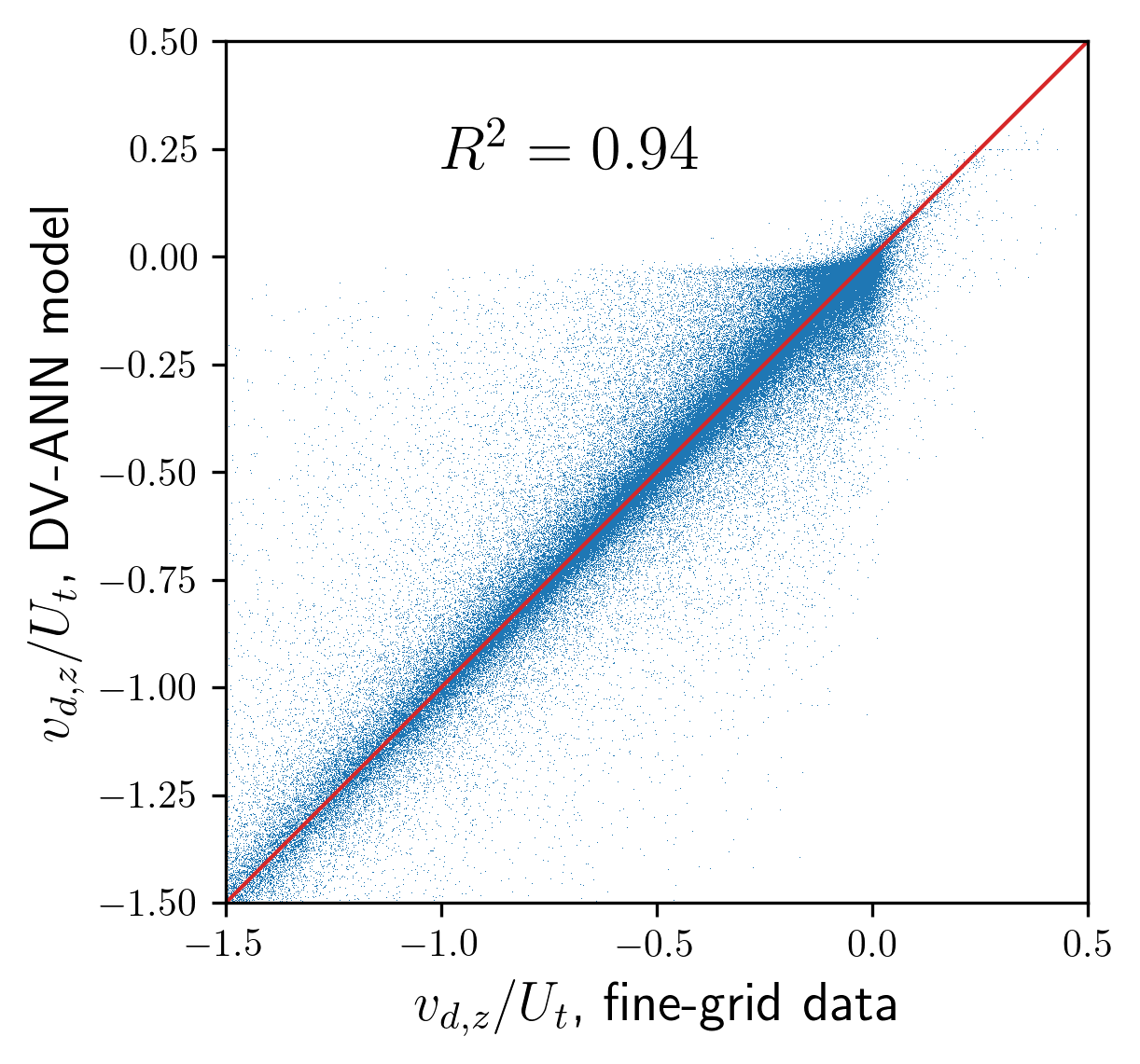}
    \end{subfigure}    
    \begin{subfigure}{0.45\textwidth}
    \includegraphics[width=\textwidth]{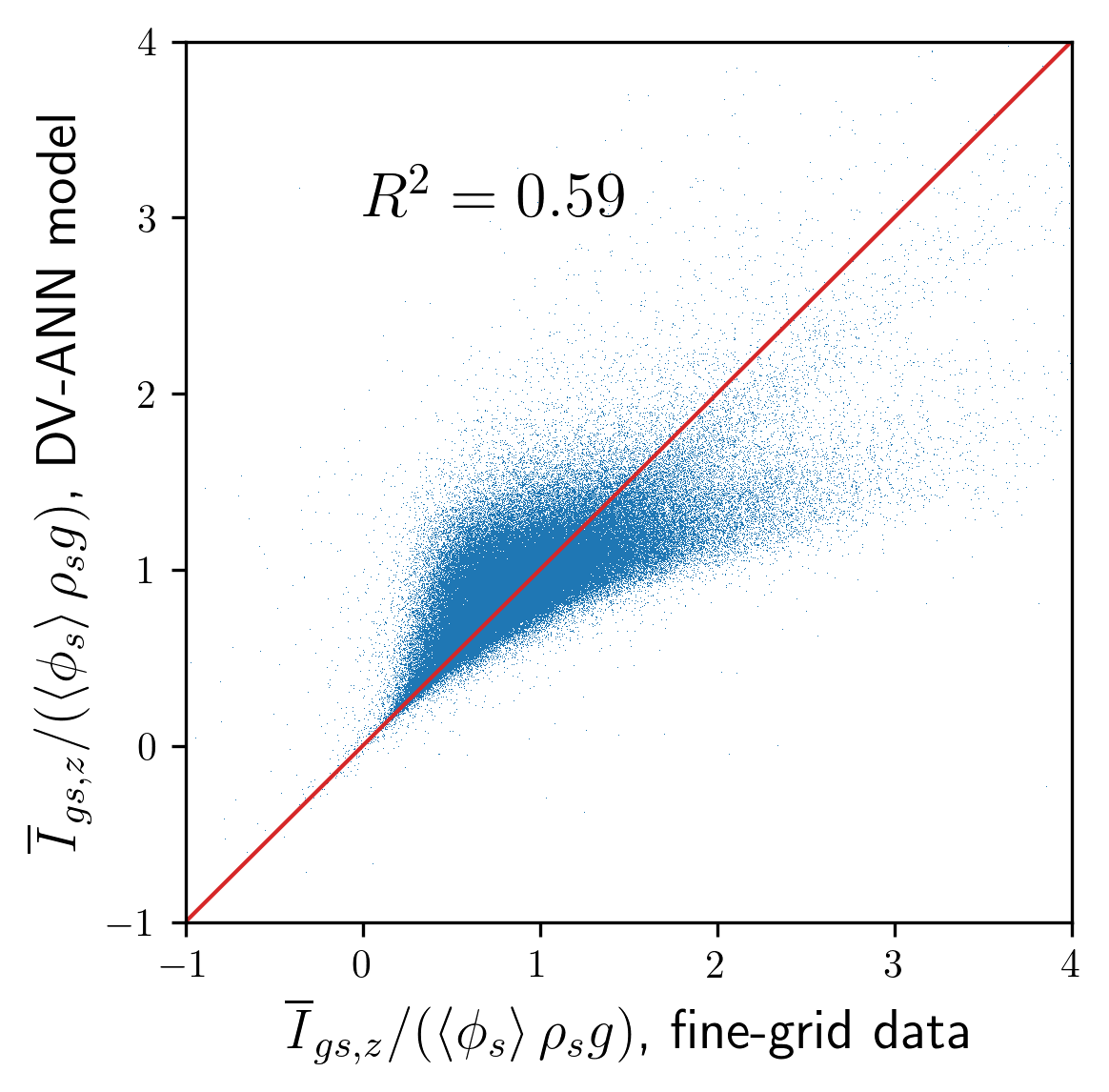}
    \end{subfigure}
    \hfill
    \begin{subfigure}{0.45\textwidth}
    \includegraphics[width=\textwidth]{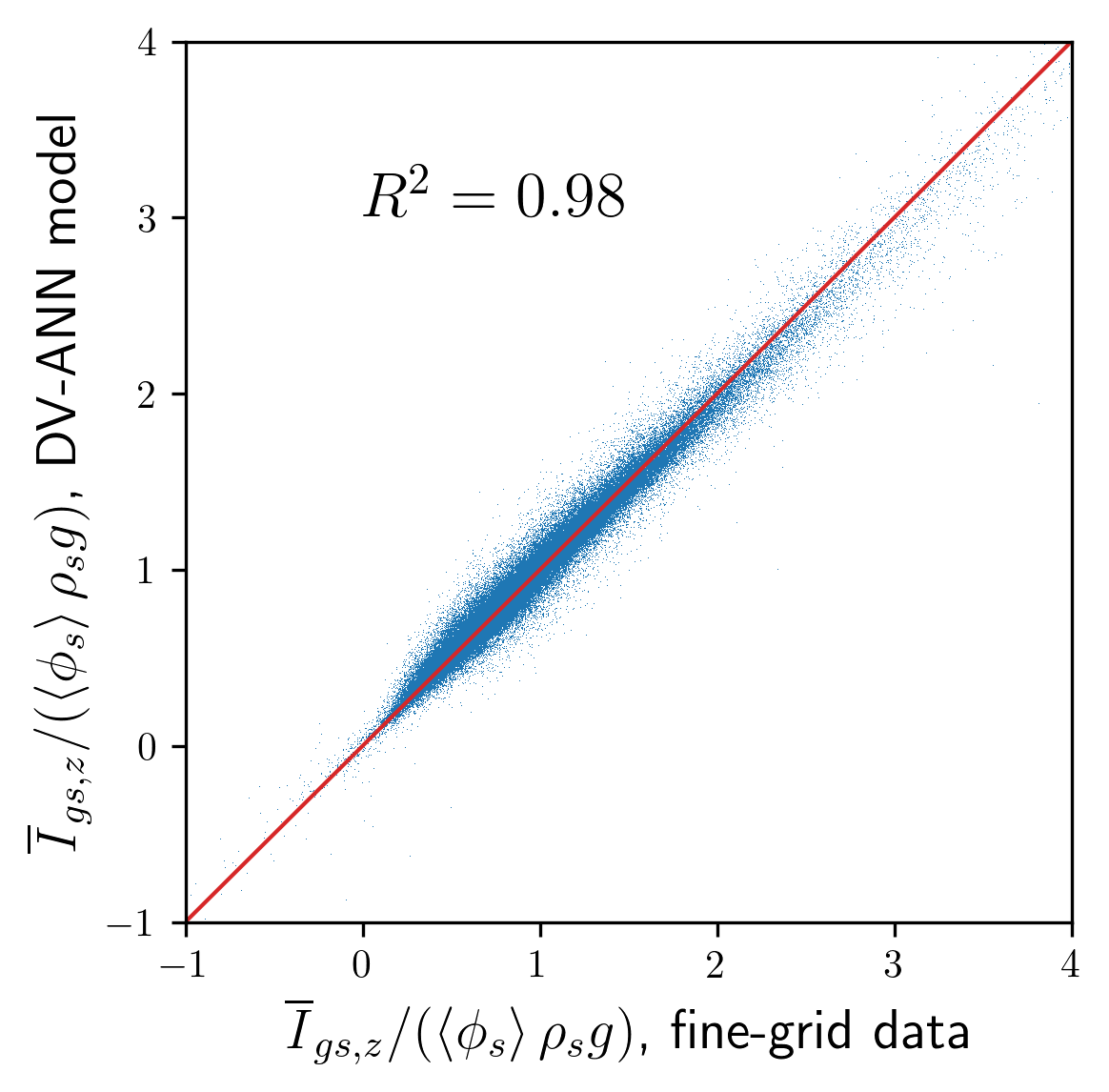}
    \end{subfigure}
    \caption{Assessment of the DF-ANN model predictions on the drift flux (top) and on the filtered drag force (bottom) using explicit Eq. \eqref{eq:drift_flux}. Left: 3-marker model $(\bar \phi_s, \bar \Delta, \tilde u_{slip,z})$. Right: 4-marker model  $(\bar \phi_s, \bar \Delta, \tilde u_{slip,z}, \frac{\partial \bar p_g}{\partial z})$.}
    \label{fig:DV-ANN-3vs4marker}
\end{figure}

\begin{figure}
\begin{subfigure}{0.45\textwidth}
    \includegraphics[width=\textwidth]{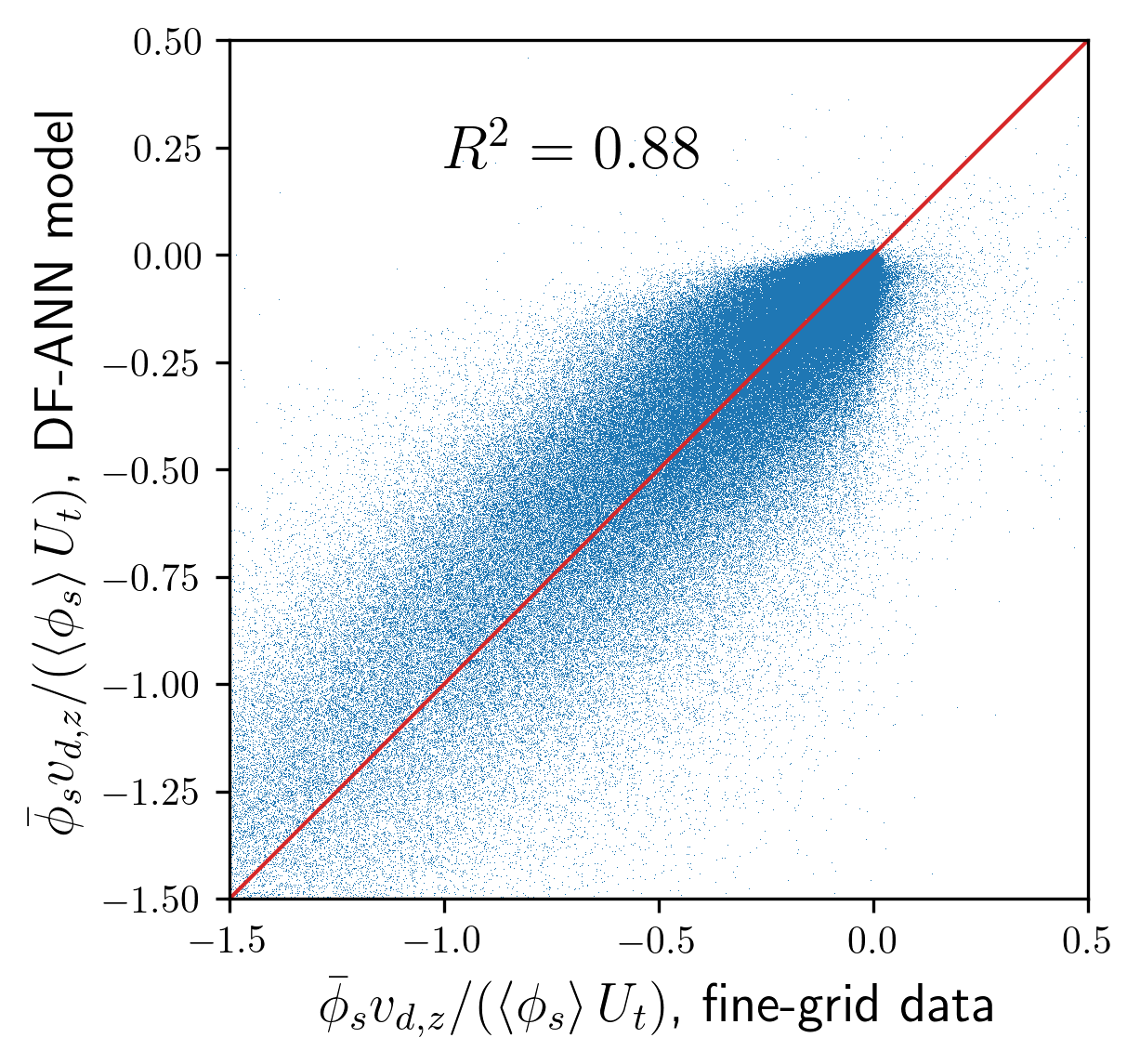}
    \end{subfigure}
    \hfill
    \begin{subfigure}{0.45\textwidth}
    \includegraphics[width=\textwidth]{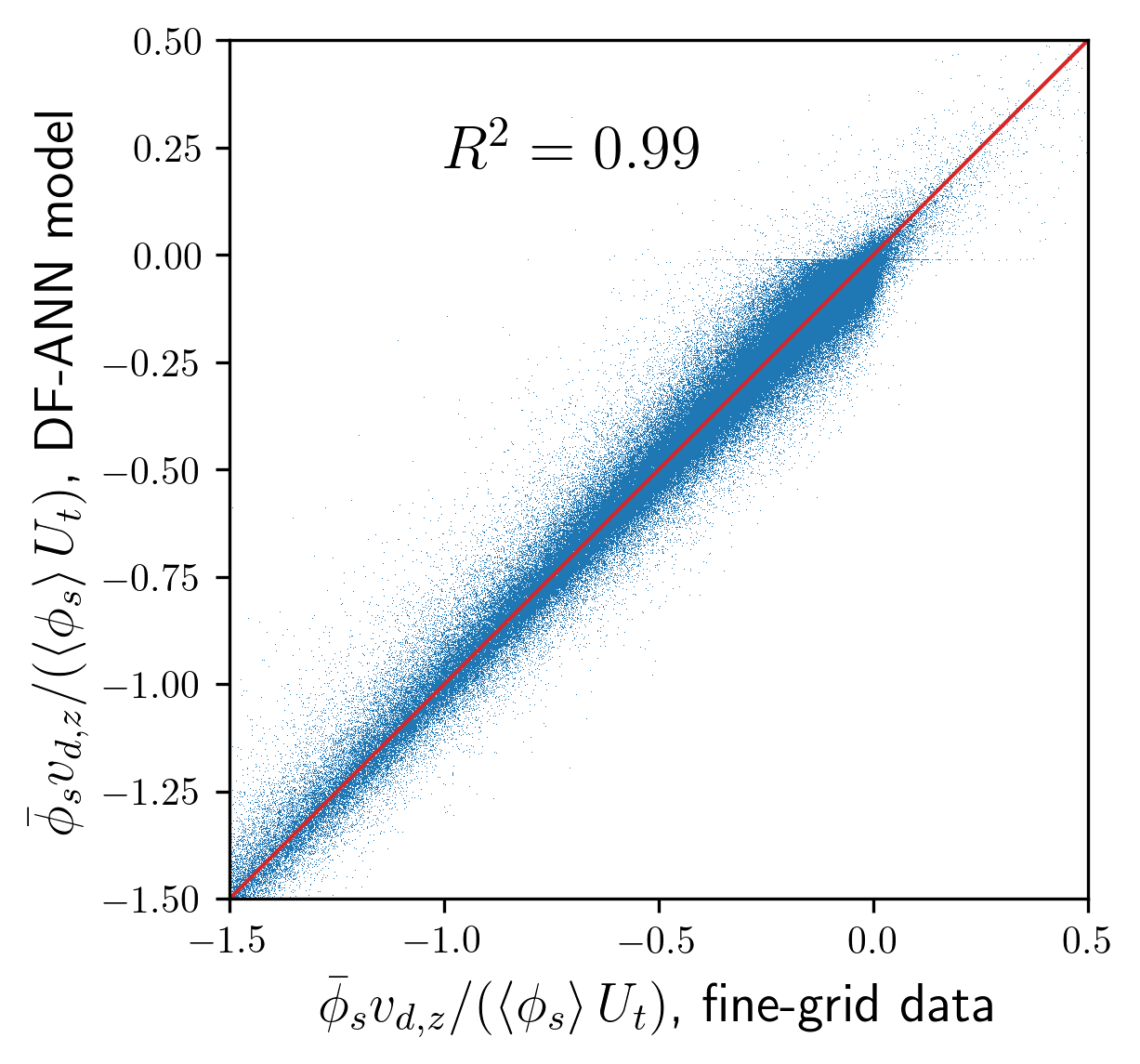}
    \end{subfigure}
    \begin{subfigure}{0.45\textwidth}
    \includegraphics[width=\textwidth]{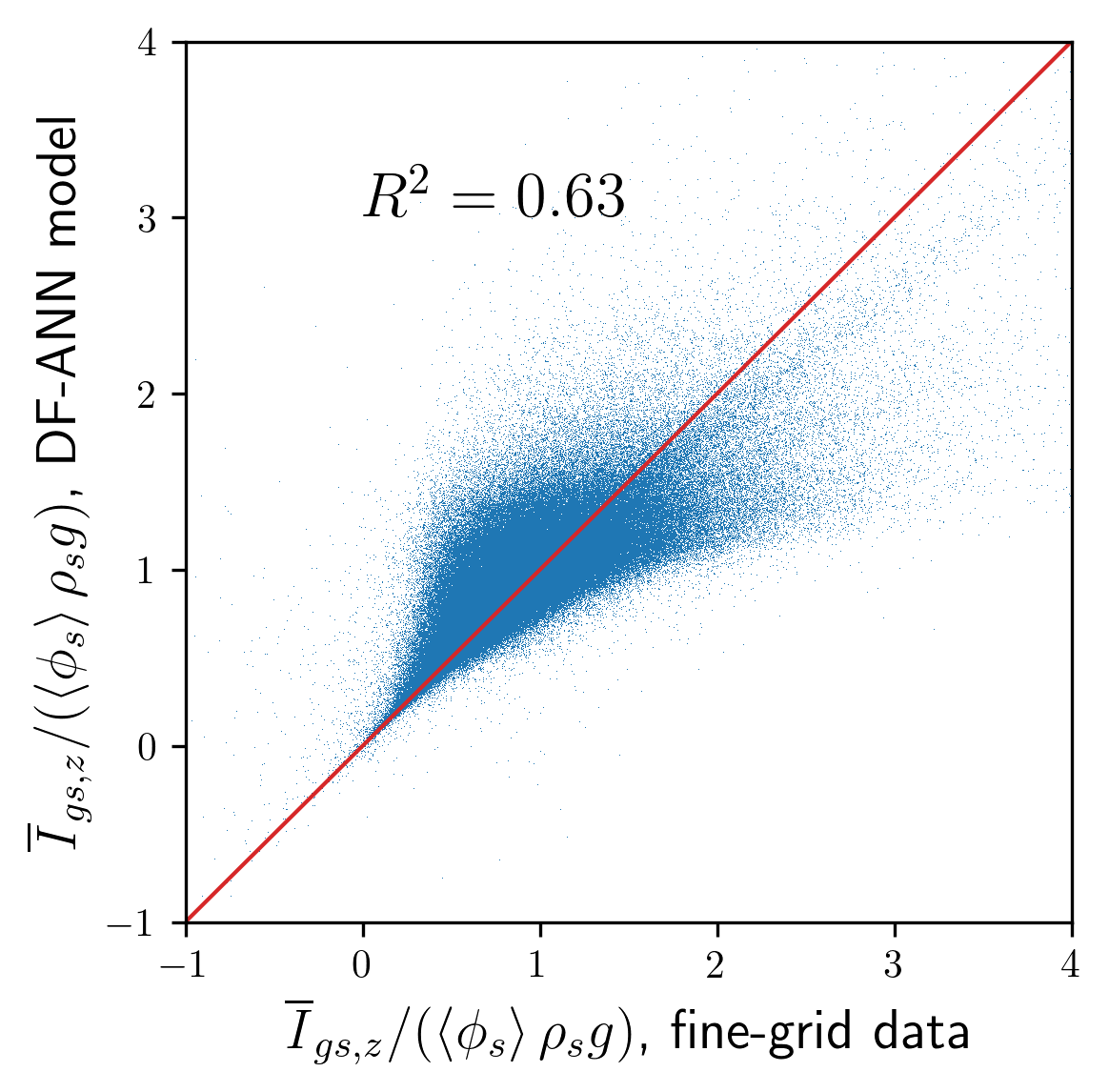}
    \end{subfigure}
    \hfill
    \begin{subfigure}{0.45\textwidth}
    \includegraphics[width=\textwidth]{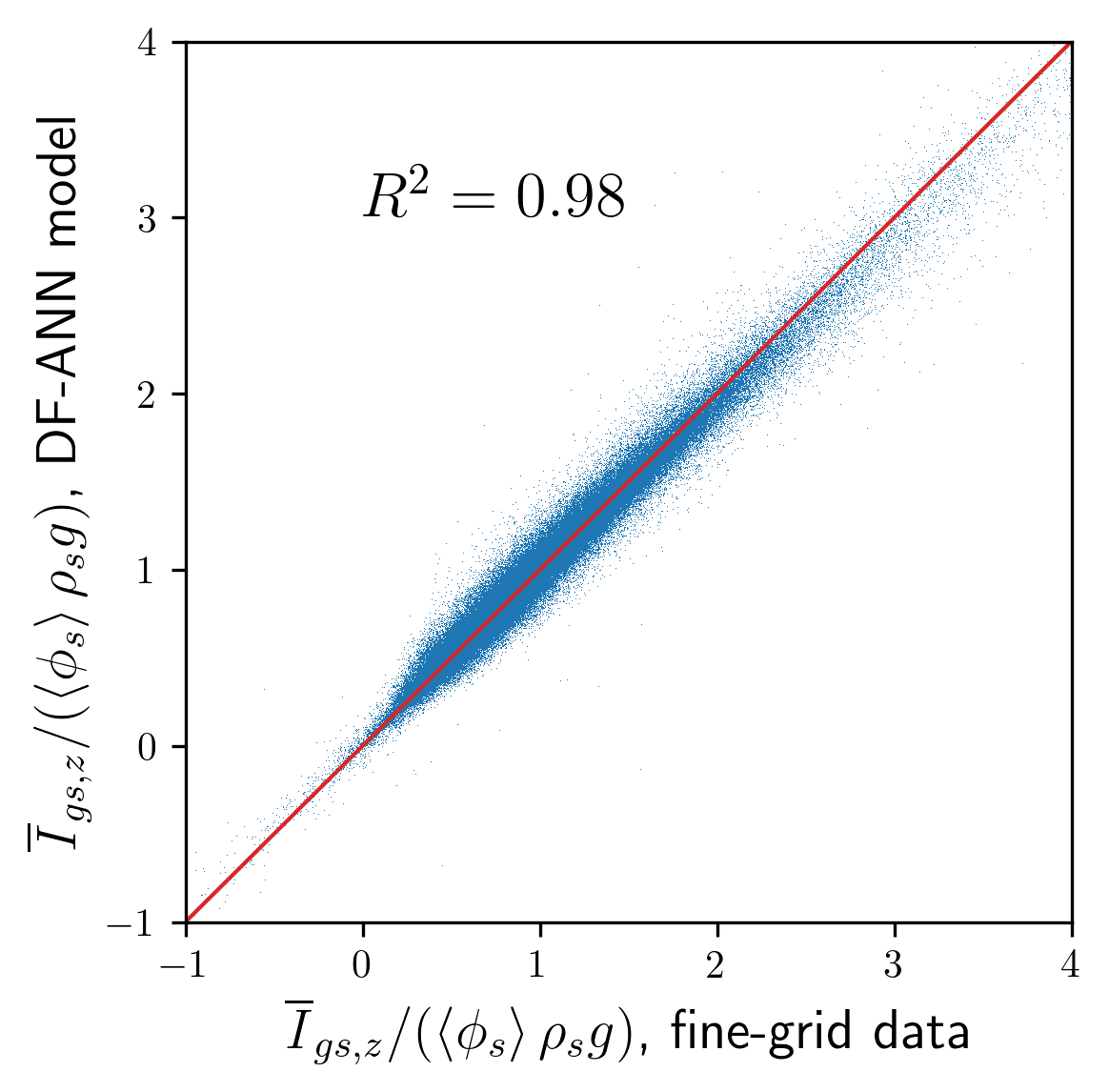}
    \end{subfigure}
    \caption{Assessment of the DF-ANN model predictions on the drift flux (top) and on the filtered drag force (bottom) using explicit Eq. \eqref{eq:drift_flux}. Left: 3-marker model $(\bar \phi_s, \bar \Delta, \tilde u_{slip,z})$. Right: 4-marker model  $(\bar \phi_s, \bar \Delta, \tilde u_{slip,z}, \frac{\partial \bar p_g}{\partial z})$.}
    \label{fig:DF-ANN-3vs4marker}
\end{figure}

We show in Figures \ref{fig:DV-ANN-3vs4marker} and \ref{fig:DF-ANN-3vs4marker} the predictive capability of the ANN model when the target quantity is the drift flux or the drift velocity, respectively. It can be observed that the vertical component of the drift flux $\bar \phi_s v_{d,z}$ is more accurately captured by the ANN than the drift velocity itself, either with the 3- or 4-marker model. However, when the result of the ANN is inserted into the explicit relations given by Eqs. \eqref{eq:filtered_drag} and \eqref{eq:filtered_drag_drift_flux} to estimate the filtered drag force, the 3-marker DV-ANN and DF-ANN models prove to be considerably inferior to the 4-marker models. Both 4-marker DV-ANN and DF-ANN models perform equally well in dilute flows; in contrast, \citet{Jiang2018} found that DF-ANN was much better than DV-ANN in the dense flow regime. Taken together, the 4-marker DF-ANN model appears to be better suited than DV-ANN and FD-ANN models for both dilute and dense flows.


\begin{figure}
    \begin{subfigure}{0.475\textwidth}
    \includegraphics[width=\textwidth]{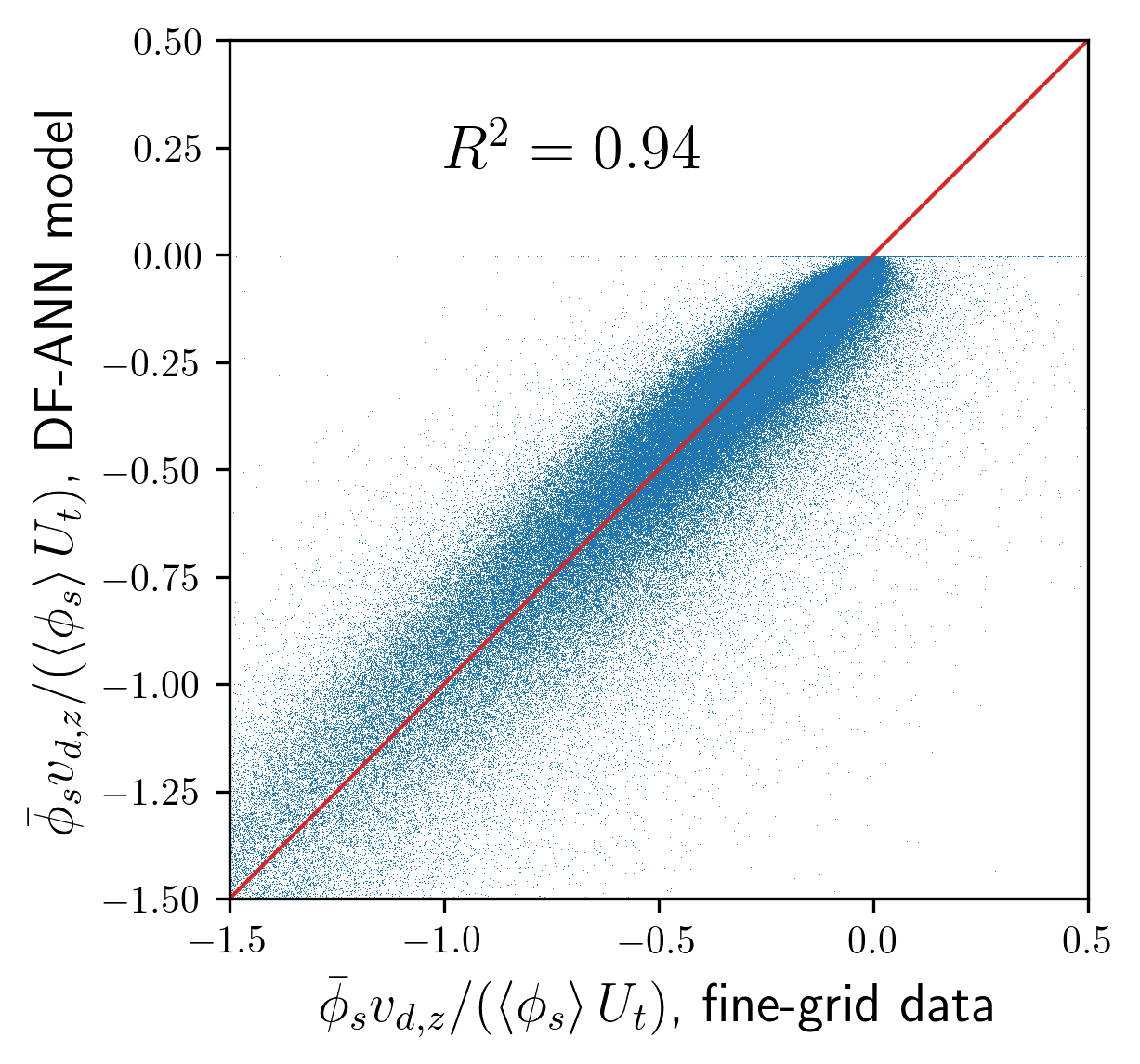}
    \end{subfigure}
    \hfill
    \begin{subfigure}{0.45\textwidth}
    \includegraphics[width=\textwidth]{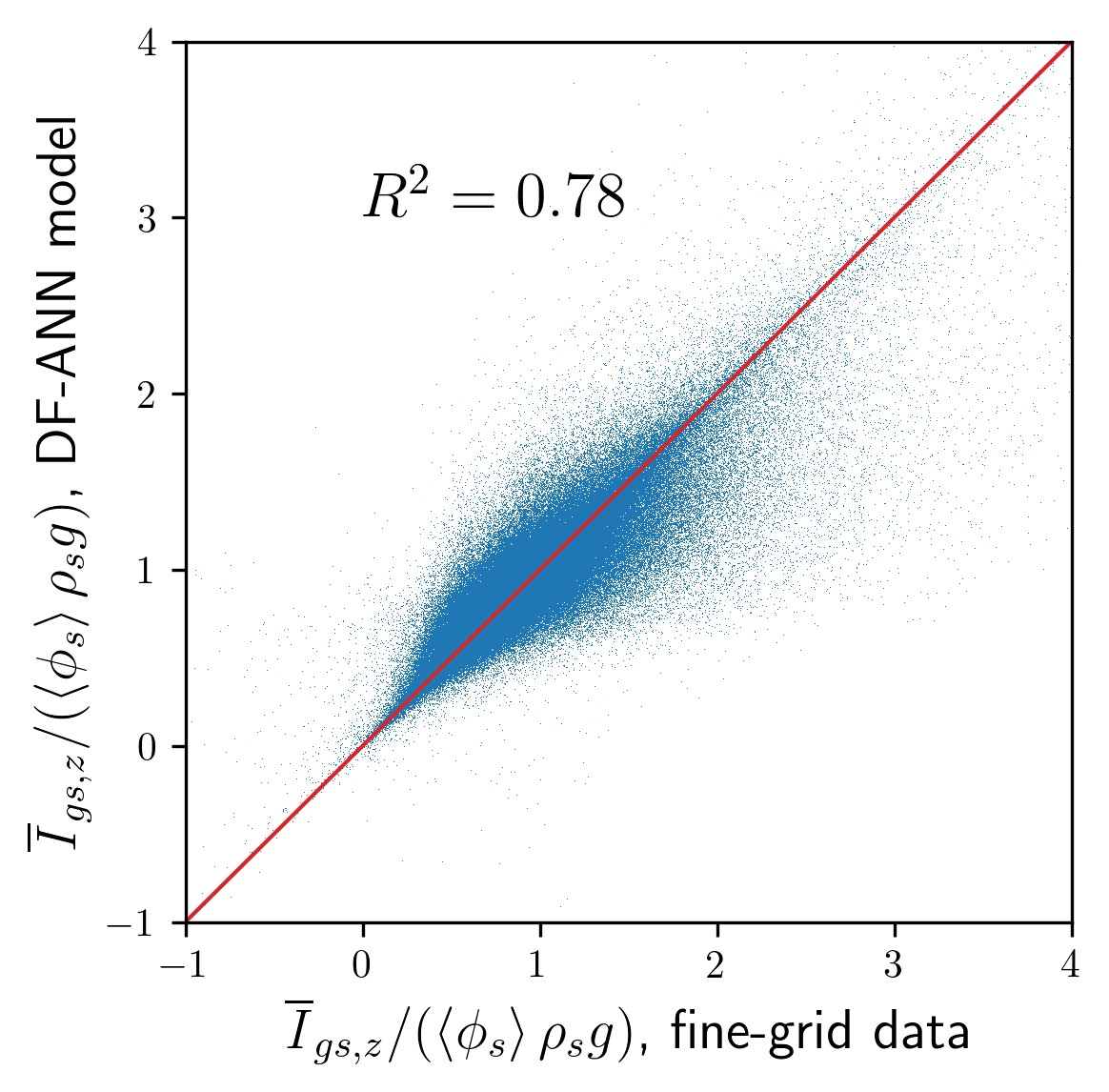}
    \end{subfigure}
    \caption{Assessment of the DF-ANN model predictions on the drift flux (left) and on the filtered drag force (right) using Eq. \eqref{eq:filtered_drag_drift_flux} with the scalar variance-based 4-marker model $(\bar \phi_s, \bar \Delta, \tilde u_{slip,z}, \overline{\phi_s^{\prime^2}})$.}
    \label{fig:DF-ANN-4markerSV}
\end{figure}

It can be inferred from Figure \ref{fig:DF-ANN-4markerSV} that the 4-marker ANN model described by 
\be
\bar \phi_s v_{d,z} = f\left(\bar \phi_s, \bar \Delta, \tilde u_{slip,z}, \overline{\phi_s^{\prime^2}} \right)
\ee
is capable of predicting the drift flux in the vertical direction with a quite high accuracy ($R^2 = 0.94$). However, the quality of the model for the filtered drag force decreases significantly ($R^2=0.78$) as previously observed with the 3-marker DF-ANN and DV-ANN models. We can therefore conclude that the vertical component of the filtered gas phase pressure gradient performs better as the fourth marker than SV for this reference case.

\citet{Zhang2020} argued that the flow information of the neighboring grid cells was crucial in predicting the local filtered drag, which is inherently provided by their CNN. Yet, if one aims to build explicit models inspired by machine learning approach, we should identify which differential quantities adds most information. To that end, we also tested 4-marker DF-ANN models where the filtered gas phase pressure gradient in the vertical direction is replaced by the vertical component of the filtered solid volume fraction gradient or the filtered solid phase velocity gradient. These attempts yielded slightly better results than those of the 3-marker model, without achieving the same predictive capacity as the gas pressure gradient-based 4-marker model. Therefore, the filtered gas phase pressure gradient appears to be the most promising fourth marker in both dense and  dilute regimes. 

\begin{figure}
    \centering
    \includegraphics[width=0.9 \textwidth]
{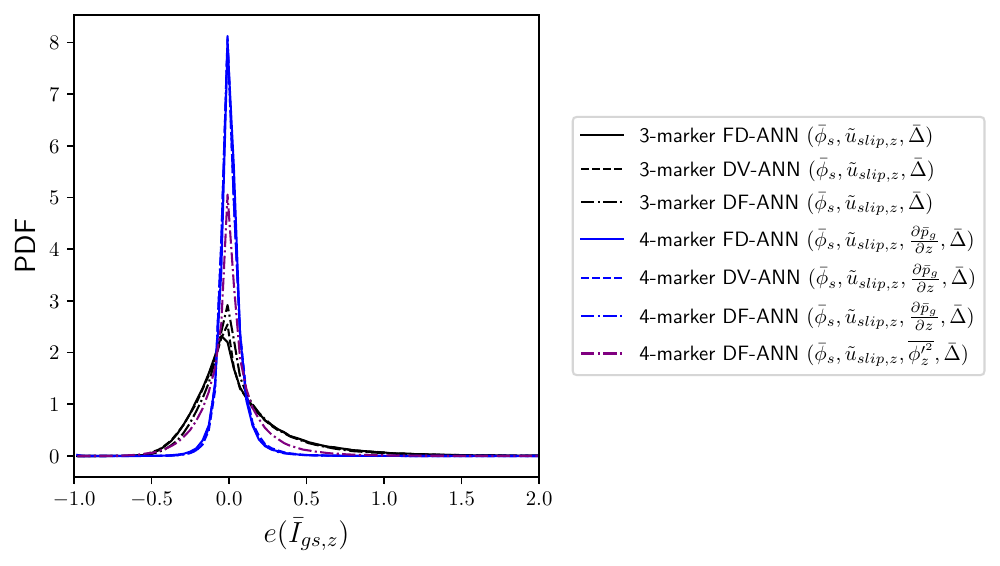}
    \caption{Relative error on the vertical component of the filtered drag force predictions for the various ANN models investigated.}
    \label{fig:error_FilteredDrag}
\end{figure}

Figure \ref{fig:error_FilteredDrag} shows the PDF of the relative error on the filtered drag force predictions. The different ANN approaches (FD, DV, DF) lead to  very similar distributions of the modeling error in the 3-marker case, although the DF-ANN model displays a slightly narrower distribution. In the 4-marker case, the PDF curves of the different ANN models are not distinguishable and are symmetric. The scalar variance-based 4-marker model leads to a narrower distribution of the error that the 3-marker model, without reaching the level of accuracy of the filtered gas phase pressure gradient-based 4-marker model, as discussed above. Based on all these considerations, we consider only the DF-ANN model for the rest of this article.

\subsubsection{Generalized ANN model}
\begin{figure}
    \centering
    \begin{subfigure}{0.48\textwidth}
    \includegraphics[width=\textwidth]{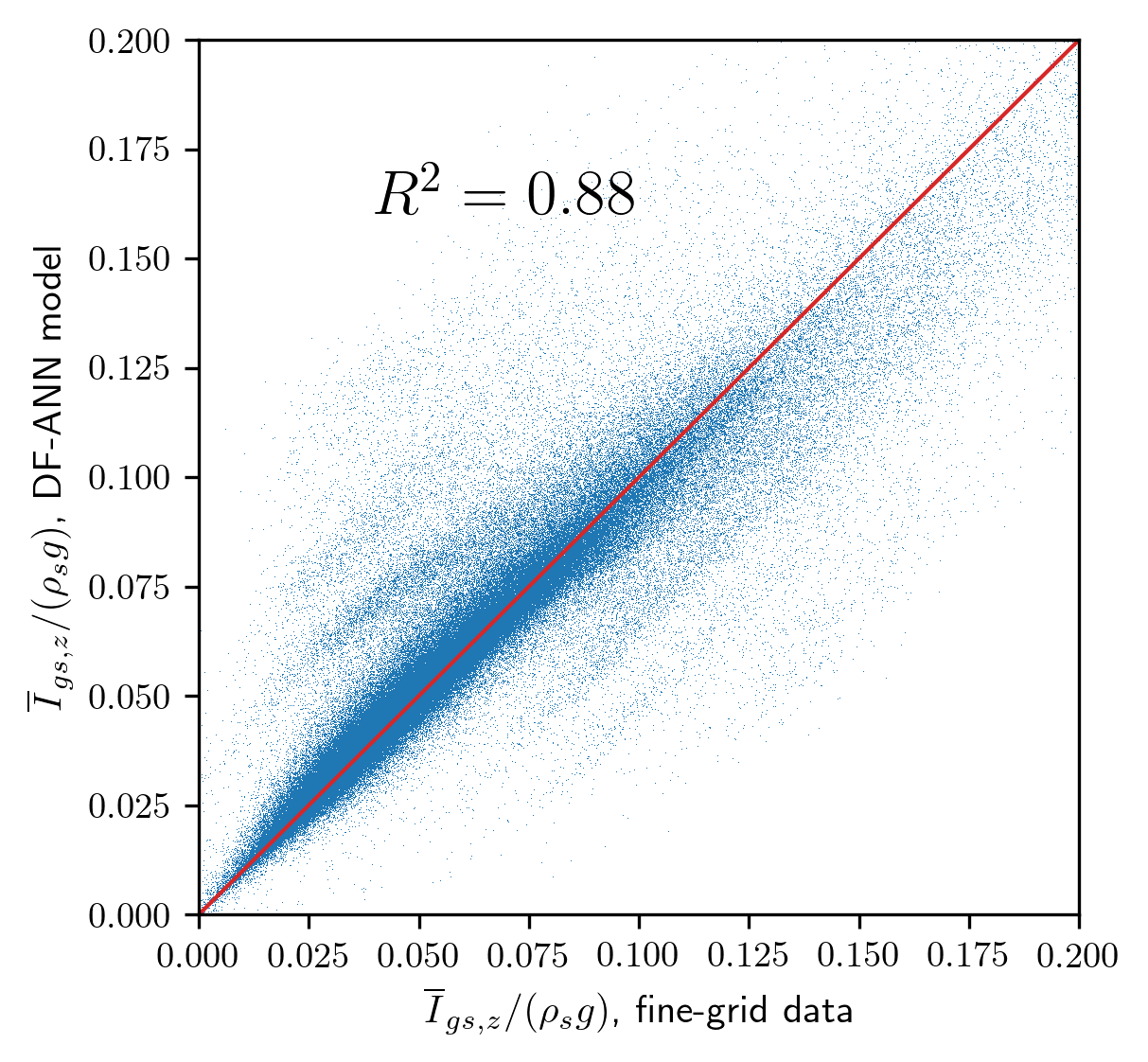}
     \end{subfigure}
    \hfill
    \begin{subfigure}{0.48\textwidth}
     \includegraphics[width=\textwidth]{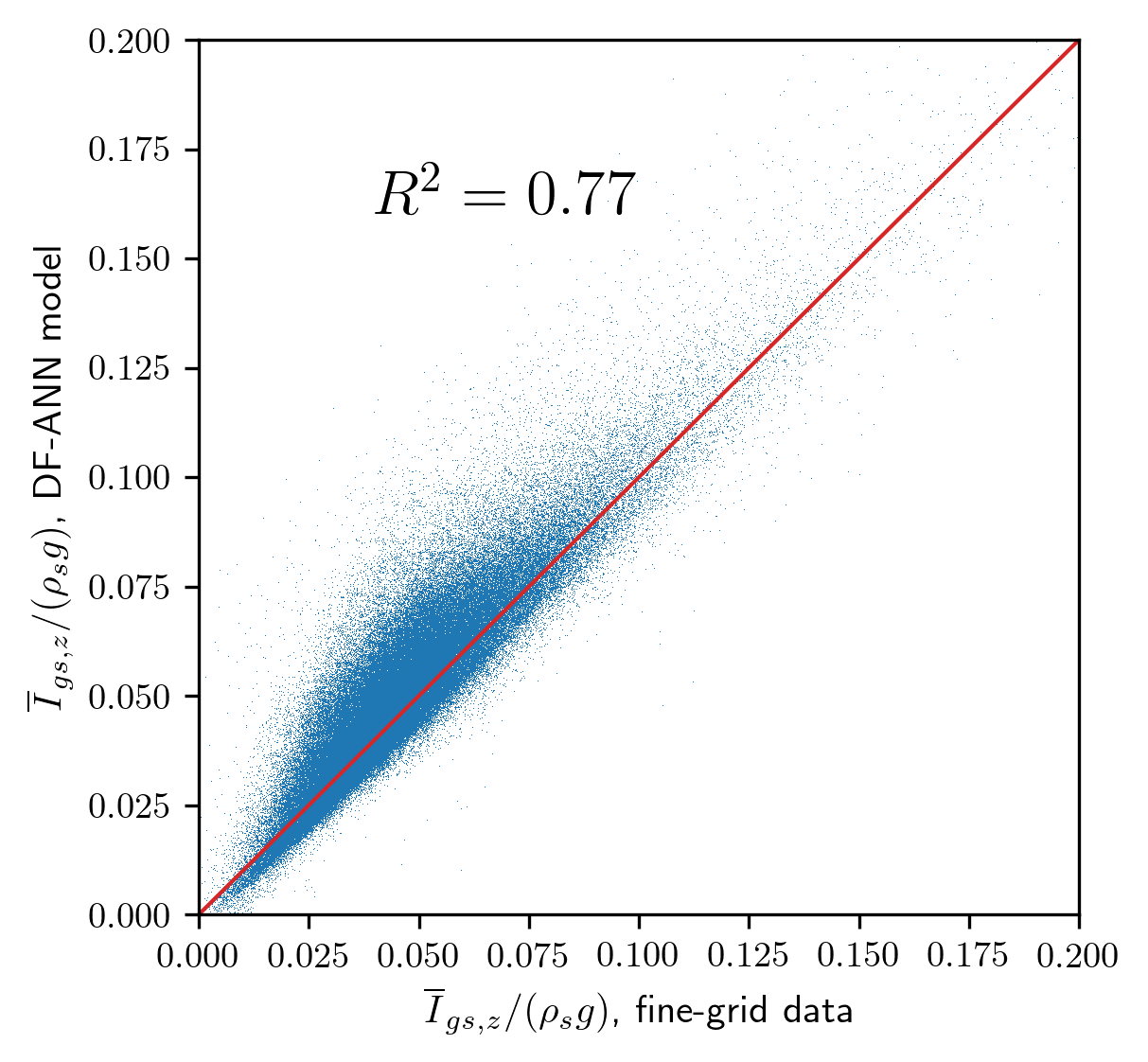}
     \end{subfigure}
     \caption{Assessment of the generalized DF-ANN model prediction with characteristic length scale $L_c = L_{c,\mathrm{I}}$: testing on Cases 1 to 8 (left) and on Case 9 (right)}
    \label{fig:model8}
\end{figure}

For a practical use in large-scale simulations, it is desirable that the filtered drag force ANN model can be generalized to a wide range of physical parameters. To that end, the input markers of the ANN must be made dimensionless using proper characteristic scales and the ANN should be trained with different datasets to cover a large range of these dimensionless input quantities.
This approach has been investigated by  \citet{Jiang2021} for bubbling to turbulent fluidized beds, but the corresponding range of solid volume fractions was confined to the fairly dense regime. To take the analysis further, simulation results from Cases 1 to 8 in Table~\ref{tab:parameters} have been used to train a generalized drift flux ANN model. Assuming that the 4-marker DF-ANN model examined in the previous section is sufficient to capture the filtered drag force for constant physical parameters, a more general DF-ANN model can expressed as 
\be
\bar \phi_s v_{d,z} = f\left(\bar \phi_s, \tilde u_{slip,z}, \bar \Delta, \frac{\partial \bar p_g}{\partial z}, d_p, \rho_s, e_c, \rho_g, \mu_g\right).
\label{eq:drift_flux_generalANN}
\ee
The restitution coefficient is kept constant in our TFM simulations, so that its influence could not be assessed. 
The scaling proposed by \citet{Jiang2021} in the dense regime to reduce the number of independent variables is given by Eq. \eqref{eq:Jiang-ANN_model}. They could not  discriminate between the three different definitions of $L_c$ usually found in the literature: $L_{c,\mathrm{I}} = d_p \Fr_p^{1/3}$ and $\mathlarger{L_{c,\mathrm{II}} = \frac{U_t^2}{g} = d_p \Fr_p}$ and $L_{c,\mathrm{III}} = d_p$, and the authors set $L_c = L_{c,\mathrm I}$. 
We start by adopting the same characteristic length in our analysis. 
Figure \ref{fig:model8} shows the predictions of the generalized DF-ANN model described by Eq. \eqref{eq:Jiang-ANN_model} with $L_c = L_{c,\mathrm I}$. The  training has been performed using 80\% of the data points from Cases 1 to 8 while the remaining 20\% have been preserved for testing purpose (left side of Figure \ref{fig:model8}). It can be observed that the generalized ANN model is able to predict the filtered drag force with decent accuracy ($R^2=0.88$). 
The model is also tested on Case 9 (right side of Figure \ref{fig:model8}), which was not used for the training process. 
This test case aims to verify that the scaling proposed in Eq. \eqref{eq:Jiang-ANN_model} is  justified in the dilute regime and that the range of parameters spanned by the training dataset (Cases 1 to 8) is sufficient to build a robust generalized model. It is shown that the scatter of the model increases slightly ($R^2 = 0.77$). 
This observation may have several causes: 
\begin{itemize}
    \item the set of dimensionless markers in Eq. \eqref{eq:Jiang-ANN_model} is imperfect or incomplete; 
    \item the range of parameters covered by the dataset formed by Cases 1 to 8 is not sufficient to train a robust, generalized DF-ANN model and additional data should be fed to the neural network; or 
    \item the neural network architecture is sub-optimal.
\end{itemize}
Though it is hard to make a definitive conclusion, our attempts to increase the complexity of the network (number of layers and number of nodes) did not yield improved predictions. By changing the definition of the characteristic length scale, using $L_c = L_{\mathrm {II}}$, we observed similar performances on Cases 1 to 8, but a dramatic decline of the model performance on Case 9. 

\begin{figure}
    \centering
    \begin{subfigure}{0.48\textwidth}
    \includegraphics[width=\textwidth]{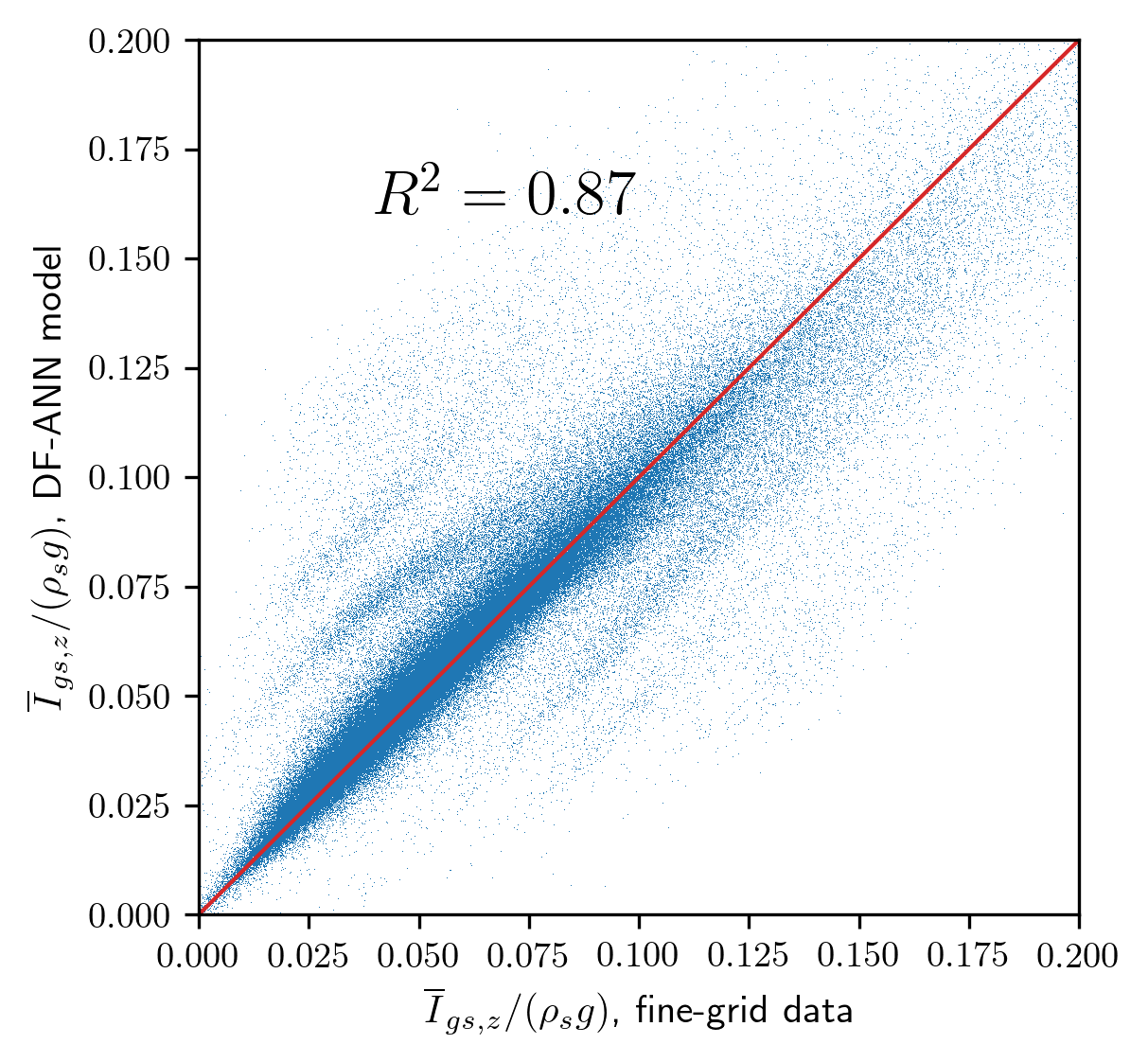}
     \end{subfigure}
    \hfill
    \begin{subfigure}{0.48\textwidth}
    \includegraphics[width=\textwidth]{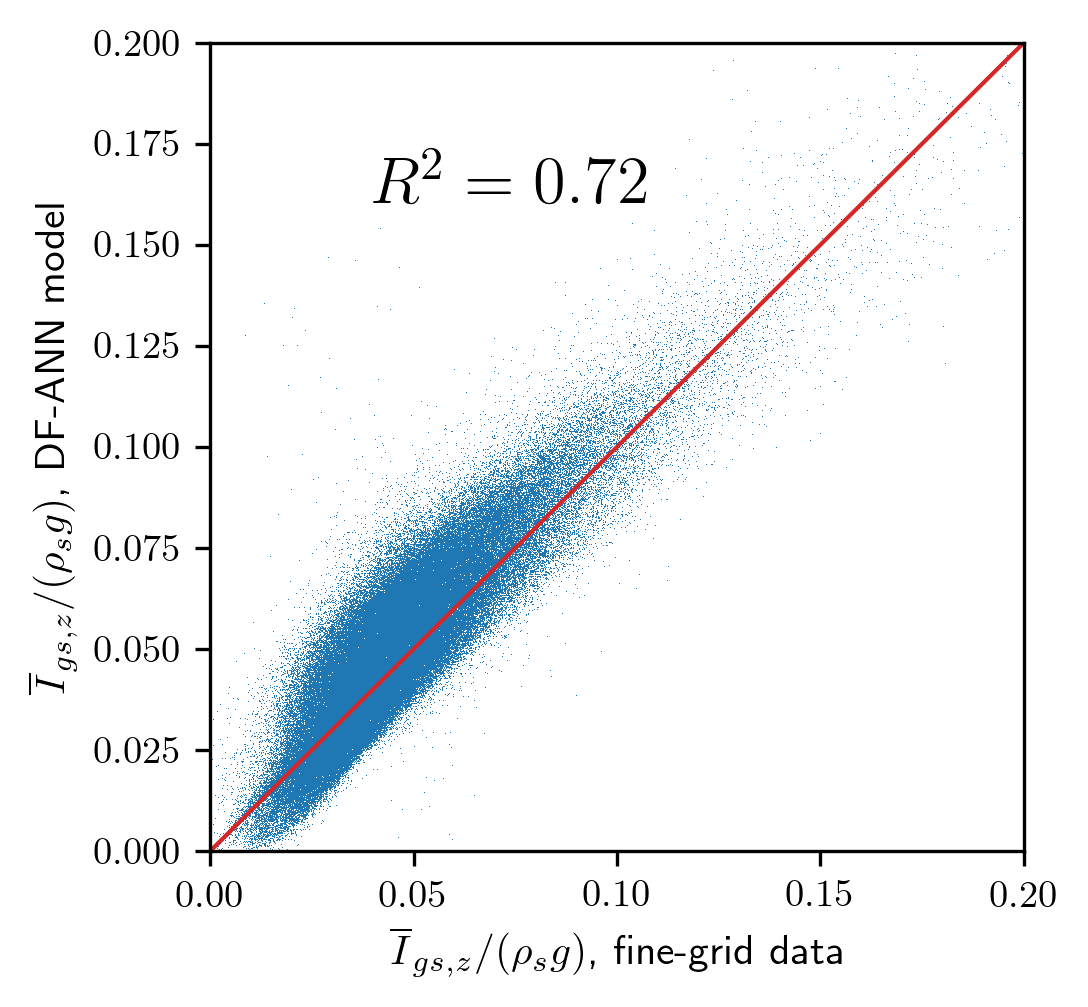}
     \end{subfigure}
     \caption{Assessment of the generalized DF-ANN model prediction with characteristic length scale $L_c = d_p$ and $\Fr_p$ added as a distinct marker}
    \label{fig:model9}
\end{figure}

Then, considering that the characteristic length scale could be more generally written as $L_c = d_p \mathcal F(\Fr_p, \Rey_p)$ (where $\mathcal F$ is some function of the Froude and Reynolds numbers), we studied the case where $L_c$ is taken equal to the particle diameter ($L_c = L_{c,\mathrm{III}}$) and the Froude number is added as an additional marker in Eq. \eqref{eq:Jiang-ANN_model}, i.e. 
\be
\frac{\bar\phi_s v_{d,z}}{\phi_{s,\mathrm{max}} U_t} = f\left(\frac{\bar \phi_s}{\phi_{s,\mathrm{max}}}, \frac{\tilde u_{slip,z}}{U_t}, \frac{\bar \Delta}{d_p}, \frac{1}{\rho_s g}\frac{\partial \bar p_g}{\partial z}, \Rey_p, \Fr_p \right).
\label{eq:general_DF-ANN_model_v2}
\ee
It is shown in Figure \ref{fig:model9} that the predictions of this model are not superior to the one presented above with the presumed definition of the characteristic length scale, and the choice $L_c = L_{c,\mathrm{I}}$ appears the best option for now, also in the dilute or moderately dense regime. Nevertheless, future studies should clarify whether a more advanced ANN architecture or a different set of dimensionless markers could improve the predictive capacity. We also note that we used 8 different cases to train our model while \citet{Jiang2021} trained their model with more than 20 different cases.  Thus, enlarging the dilute regime datasets in future studies to include more cases  spanning a wider range of parameters appears to be the best approach to further improve the model's predictive capabilities. It is reassuring to know (based on the present study) that DF-ANN model for drag correction trained with a comprehensive set of data can bridge both dilute and dense flow conditions. 

\clearpage
\section{Neural Network Modeling of Filtered Solid Stresses}
\label{section:filtered_stresses_ANN}

Machine learning based modeling of the filtered solid phase stresses has also been addressed in the literature, though more sparsely. By relying on an eddy-viscosity concept, Ouyang et al. \cite{Ouyang2021} used fine-grid 2-D simulations results to train a 7-layer ANN and predict the filtered solid phase pressure and the effective viscosity of the solid phase. These authors claim that an anisotropic model using the filtered solid volume fraction, the filter size, the filtered solid phase velocity and its gradient improves the accuracy of the predictions with respect to isotropic markers, e.g. using only the norm of the rate-of-deformation tensor. 
Recently, \citet{Ouyang2022} implemented their ANN solid stress model into a fTFM solver and compared their results with explicit closures. They conclude that anisotropic models are needed for low fluidization velocities in a laboratory-scale dense gas–solid fluidized bed.
In this study, we adapt the advanced ANN architecture proposed by  \citet{ling2016reynolds} to model the Reynolds stresses in single phase flow turbulence with embedded Galilean invariance, and formulate a model for the subgrid solid phase stress in fTFM.  

\subsection{Artificial Neural Network Architecture for Solid Phase Subgrid Stresses}

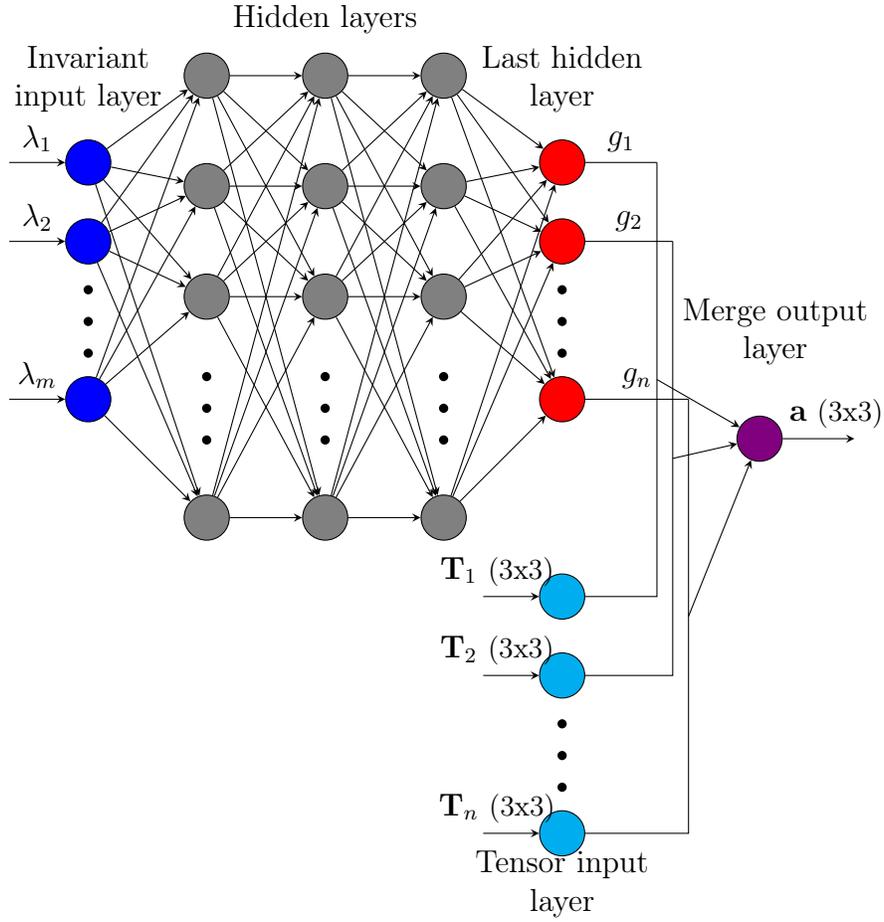
\begin{figure}
    \centering
    \tikzset{every picture/.style={scale=0.7}}
    \input{TBNN_stresses_drawing}
    \caption{Tensor Basis Neural Network (TBNN) architecture for modeling of the filtered solid phase stresses. $\lambda_1$,... $\lambda_5$ refer to the scalar bases, given by Eq.~\eqref{eq:scalar_basis_tbnn}, and $\mathbf T^{(1)}$...$\mathbf T^{(10)}$ refer to the tensor bases, which are functions of the strain- and rotation-rate tensors given by Eq.~\eqref{eq:tensor_basis_tbnn}. }
    \label{fig:FD_TBNN_arch}
\end{figure}

To the best of the authors knowledge, only eddy viscosity-type ANN models inspired by Smagorinsky \cite{smagorinsky1963} model in single phase flow turbulence have been developed to close the deviatoric part of the filtered solid stresses. 
The filtered solid stress tensor is split into its isotropic and deviatoric parts as
\be
\boldsymbol \sigma_{s} = \boldsymbol \tau_{s} + \frac{1}{3} \mathrm{tr}(\boldsymbol \sigma_{s})\mathbf I. 
\label{eq:sigma_s}
\ee
The filtered solid pressure, or so-called meso-scale pressure, is given by 
\be
P_{s,\mathrm{meso}} = \frac{1}{3} \mathrm{tr}(\boldsymbol \sigma_{s}). 
\label{eq:P_s}
\ee
Smagorinsky-type models assume the alignment between the deviatoric solid stress tensor and the deviatoric part of the filtered solid rate-of-deformation tensor, namely
\be
\boldsymbol \tau_s = 2 \mu_{s,\mathrm{meso}}  \boldsymbol{\tilde S}_s,  
\label{eq:eddy_viscosity_model}
\ee
where $\mu_{s,\mathrm{meso}}$ is the so-called meso-scale viscosity, 
which is usually estimated from filtered fine-grid data as
\be
\mu_{s,\mathrm{meso}} = \frac{\sqrt{\boldsymbol \tau_s:\boldsymbol \tau_s}}{2 \sqrt{\boldsymbol{\tilde S}_s:\boldsymbol{\tilde S}_s}}.
\ee
So far, ML-based models for the filtered solid stresses sought to describe the meso-scale pressure and meso-scale viscosity through functional models or with distinct ANNs. In single phase turbulence, the effective viscosity models have known drawbacks: they are completely dissipative, which means that they do not unveil energy backscattering \cite{dabbagh2022anisotropy}, and they do not capture accurately anisotropic stresses, even in simple shear flows~\cite{tracey2013}. In the Reynolds-averaged Navier–Stokes approach for single phase flows, various ML methods have been recently developed to model the individual components of the turbulent stresses~\cite{tracey2013, tracey2015, zhang2015}. One of the most promising ideas has been proposed by~\citet{ling2016reynolds} with a special neural network architecture referred to as a Tensor Basis Neural Network (TBNN). This architecture, sketched in Figure~\ref{fig:FD_TBNN_arch}, satisfies the Galilean invariance by relying on the decomposition of the deviatoric stress tensor as a function of a basis of tensors. In this way, the output of the neural network is not modified by an arbitrary rotation of the reference frame, which is a key principle in turbulence models development. 

The TBNN approach developed by \citet{ling2016reynolds} was inspired by the work of \citet{pope1975}, who showed that, in the single phase incompressible case, a general eddy viscosity model that is a function of the rate-of-deformation and rate-of-rotation tensors only could be expressed as a linear combination of 10 basis tensors: 
\be
\mathbf a = \sum_{n=1}^{10} g^{(n)} \left(\lambda_1, ..., \lambda_5\right) \mathbf T^{(n)}, 
\label{eq:tbnn}
\ee
where $\mathbf a$ is the deviatoric stress tensor normalized by the turbulent kinetic energy, while $\lambda_1$,... $\lambda_5$ (the scalar basis) and $\mathbf T^{(1)}$...$\mathbf T^{(10)}$ (the tensor basis) are functions of the filtered strain- and rotation-rate tensors. 
As mentioned by \citet{ling2016reynolds}, any tensor who can be expressed as Eq. \eqref{eq:tbnn} will satisfy Galilean invariance by construction. The simple eddy-viscosity model is recovered by limiting to $n=1$. 
We adopt the same approach for the modeling of the filtered solid phase stresses in gas-solid flows, where, by analogy with \citet{pope1975}, the tensor and scalar bases can be expressed as: 
\be
\begin{array}{ll}
   \mathbf T^{(1)} = \Ss \qquad & \mathbf T^{(6)} = \RR\Ss + \Ss \RR - \frac{2}{3}\mathrm{tr}(\Ss\RR) \mathbf I \\
   \mathbf T^{(2)} = \Ss\Rs - \Rs\Ss \qquad & \mathbf T^{(7)} =  \Rs\Ss\RR - \RR\Ss \Rs \\ 
   \mathbf T^{(3)} = \SS - \frac{1}{3}\mathrm{tr}(\SS) \mathbf I  \qquad& \mathbf T^{(8)} =  \Ss\Rs\SS - \SS\Rs\Ss \\ 
   \mathbf T^{(4)} = \RR - \frac{1}{3}\mathrm{tr}(\RR) \mathbf I  \qquad& \mathbf T^{(9)} =  \RR\SS - \SS\RR - \frac{2}{3}\mathrm{tr}(\SS\RR) \mathbf I \\ 
  \mathbf T^{(5)} = \Rs\SS - \SS\Rs  \qquad& \mathbf T^{(10)} = \Rs\SS\RR - \RR\SS\Rs  
\end{array}
\label{eq:tensor_basis_tbnn}
\ee
and
\be
\lambda_1 = \tr(\SS), \, \lambda_2 = \tr(\RR), \, \lambda_3 = \tr(\SSS), \, \lambda_4 = \tr(\RR \Ss), \, \lambda_5 = \tr(\RR \SS),
\label{eq:scalar_basis_tbnn}
\ee
where $\Ss$ and $\Rs$ are the suitably scaled (see below) rate-of-deformation and rate-of-rotation tensors, respectively defined by Eqs. \eqref{eq:strain_rate} and \eqref{eq:rotation_rate}.

Every tensor in the  basis defined by Eq. \eqref{eq:tensor_basis_tbnn} is traceless and symmetric, consistent with the tensor to be modeled.
The goal of the TBNN is therefore to capture the scalar functions $g^{(n)}$ in Eq. \eqref{eq:tbnn} and the construction of the final tensor is performed by the  output merge layer as shown in Figure \ref{fig:FD_TBNN_arch}.
Similar to \citet{ling2016reynolds} and \citet{pope1975}, the tensor $\mathbf a$ is  identified to the deviatoric part of the filtered solid stress scaled by the meso-scale pressure (or, equivalently, the sub-grid turbulent kinetic energy): 
\be
\mathbf a = \frac{\boldsymbol \tau_s}{3 P_{s,\mathrm{meso}}} \triangleq \boldsymbol \tau_s^*.
\label{eq:a_def}
\ee
The input tensors $\Ss$ and $\Rs$ are scaled using the time-scale $\mathlarger{\frac{U_t}{g}}$. 
The scaled deviatoric part of the solid stress tensor and its eigenvalues ($\xi_1 > \xi_2 > \xi_3$) must satisfy the following realizability conditions \cite{Banerjee2007}: 
\be
\begin{array}{ll}
   -\frac{1}{3} \leq a_{ii} \leq \frac{2}{3} & \xi_{1} \leq (3 |\xi_2|- \xi_2)/2  \\
    -\frac{1}{2} \leq a_{ij} \leq \frac{1}{2} \; \mathrm{ for } \; j\neq i & \xi_{1} \leq \frac{1}{3} - \xi_2. 
\end{array}
\label{eq:realizability_cond}
\ee
\citet{ling2016reynolds} suggest to add a post-processing step after the TBNN model to iteratively enforce conditions given by Eq. \eqref{eq:realizability_cond}. Beside the TBNN model for the deviatoric part of the stress, a distinct ANN model is needed to predict the meso-scale pressure. Once $P_{s,\mathrm{meso}}$ is known, the anisotropic stress $\boldsymbol  \tau_s$ can be retrieved from Eq. \eqref{eq:a_def} and the full stress tensor $\boldsymbol \sigma_s$ (given by Eq. \eqref{eq:sigma_s}) can be closed.
In addition to the scalar basis defined by Eq. \eqref{eq:scalar_basis_tbnn}, other scalar inputs specific to the modeling of filtered solid stresses might enter the network, namely the filtered solid volume fraction and/or the filtered slip velocity. The filter width $\bar \Delta$ is also added as an extra marker to account for the variation of the mesh size. In the present study, we only consider the reference case (Case 1) to examine the merits of TBNN-based stress modeling for fTFM analysis. Additional markers accounting for variation in physical properties (such as $\Rey_p$, $\Fr_p$) will enter the network (as was done for the drift flux model development) when the model is generalized to cover a wide a range of gas-particle systems.

\section{\textit{A Priori} Benchmark Results on ANN Solid Subgrid Stress Modeling}

\subsubsection{MLP Models for Meso-Scale Solid Pressure and Effective Viscosity}
As a preliminary step, and for a point of comparison, simple MLP models similar to the one used for the filtered drag force are built to predict the meso-scale viscosity and meso-scale pressure. In both cases, the 3-layer network architecture detailed earlier is employed with the same number of nodes (124, 32, 8). From our experience, deeper networks did not yield superior results. 
The meso-scale pressure and viscosity are scaled as follows: 
\be
\mu_{s,\mathrm{meso}}^* = \frac{\mu_{s,\mathrm{meso}}g}{\rho_s U_t^2} \hspace{1cm},  
P_{s,\mathrm{meso}}^* = \frac{P_{s,\mathrm{meso}}}{\rho_s U_t^2}.
\ee

\begin{figure}
    \centering
    \begin{subfigure}{0.48\textwidth}
    \includegraphics[width=\textwidth]{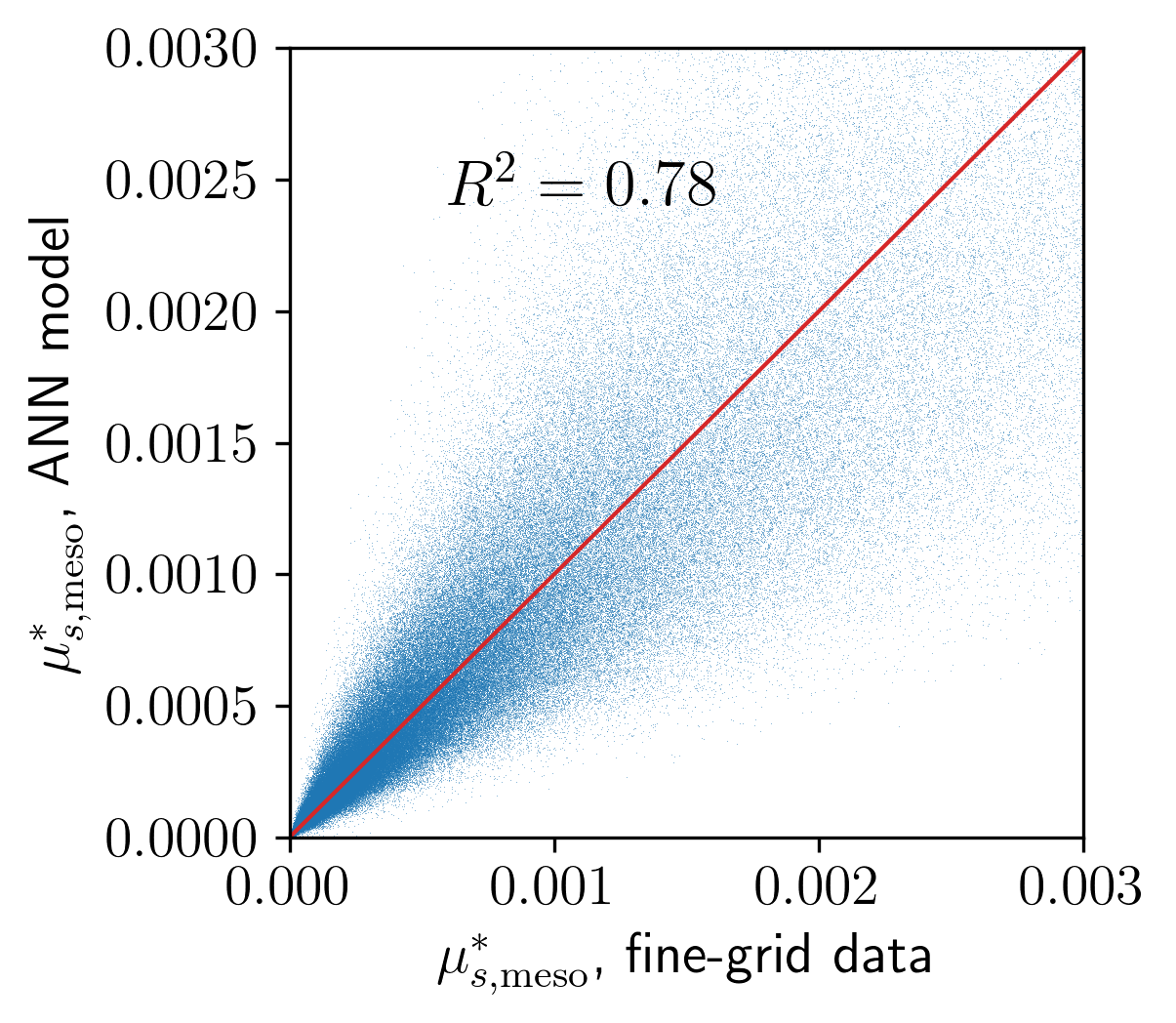}
     \end{subfigure}
    \hfill
    \begin{subfigure}{0.48\textwidth}
      \includegraphics[width=\textwidth]{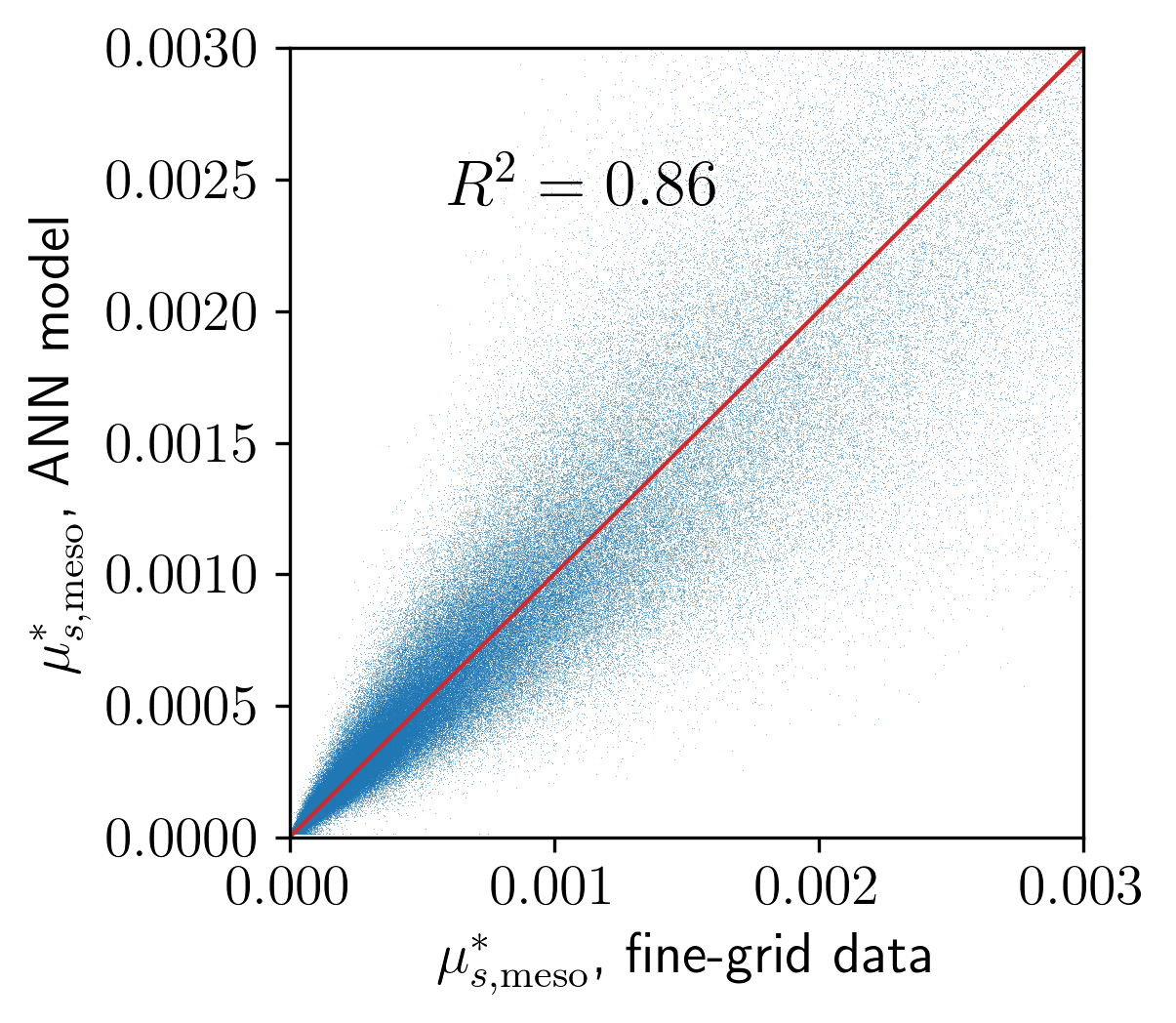}
     \end{subfigure}
     \caption{Prediction of the scaled meso-scale solid phase viscosity by the 3-marker (left) and 14-marker (right) ANN models for Case 1.}
    \label{fig:mu_mesoscale}
\end{figure}

\begin{figure}
    \centering
    \begin{subfigure}{0.48\textwidth}
    \includegraphics[width=\textwidth]{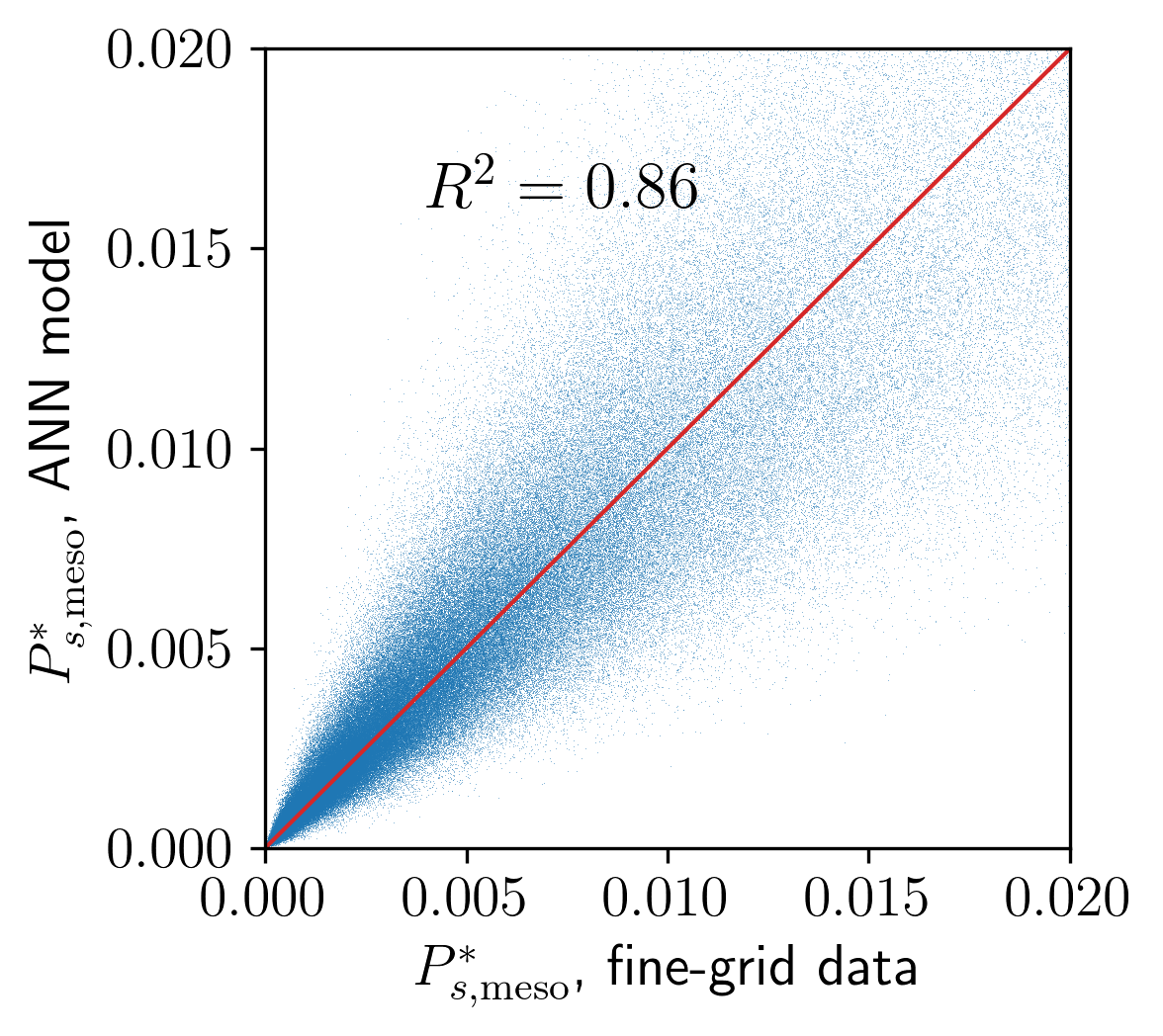}
     \end{subfigure}
    \hfill
    \begin{subfigure}{0.48\textwidth}
      \includegraphics[width=\textwidth]{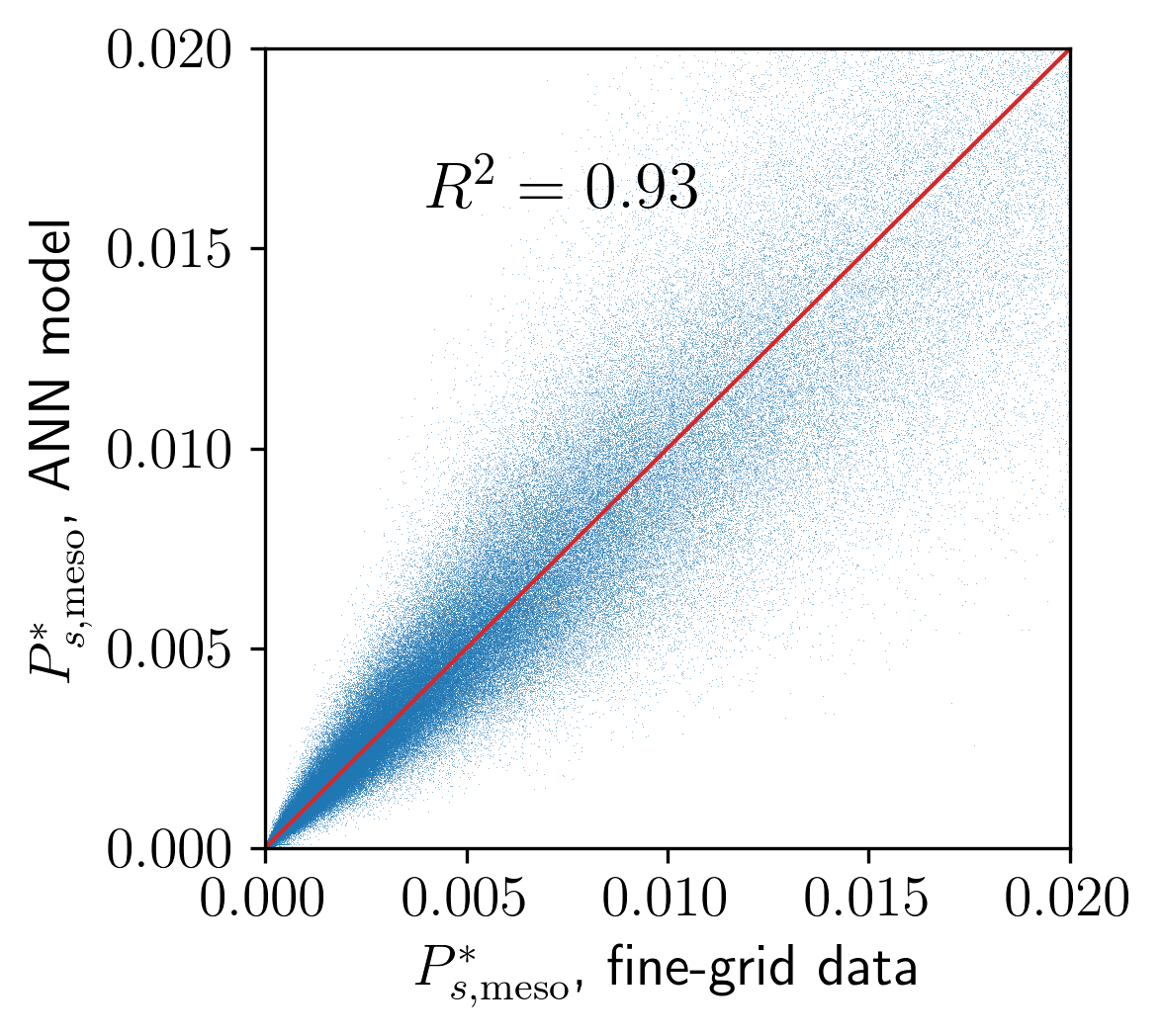}
     \end{subfigure}
      \caption{Prediction of the scaled meso-scale solid pressure by the 3-marker (left) and 14-marker (right) ANN models for Case 1.}
    \label{fig:Ps_mesoscale}
\end{figure}

In line with the work of \citet{Ouyang2021}, we compare two ANN models for $\mu_{s,\mathrm{meso}}^*$ and $P_{s,\mathrm{meso}}^*$ with different input markers: 
\begin{itemize}
    \item A 3-marker model with $\lambda_1 = \mathrm{tr}(\SS)$, $\bar \phi_s$ and $\bar \Delta$ as inputs,
    \item A 14-marker model with $\bar \phi_s$, $\bar \Delta$ and the components of $\tilde \bu_s$ and $\nabla \tilde \bu_s$ as inputs. 
\end{itemize}
\citet{Ouyang2021} concluded from their 2-D analysis that the second "anisotropic" model showed improved predictive capacity with respect to the first isotropic version. It must be stressed that this model is anisotropic in the sense that the input markers contain directional information, but the final eddy-viscosity model still relies on the alignment between $\boldsymbol \tau_s$ and $\Ss$ (see Eq. \eqref{eq:eddy_viscosity_model}). 
Nevertheless, Figures \ref{fig:mu_mesoscale} and \ref{fig:Ps_mesoscale} confirm that the more complete 14-marker ANN model fed with the individual components of $\tilde \bu_s$ and $\nabla \tilde \bu_s$ shows reduced scatter with respect to the simpler 3-marker one, both for the meso-scale pressure and viscosity. 

\subsubsection{TBNN model for the subgrid stresses}

It was shown in previous section that the meso-scale pressure and viscosity can be quite successfully captured with simple ANN models, which can be enough if one is only interested in low-order modeling of the stresses. We now address the TBNN model described earlier when a more complete description of the stresses accounting for anisotropic effects is sought.
 The TBNN network sketched in Figure \ref{fig:FD_TBNN_arch} was built with 8 hidden layers of 30 nodes. 
The activation function of the hidden layers is the ReLU and the loss function is the minimum absolute error (MAE) as for the classical MLP studied above. The MAE is  computed on all components of the final deviatoric stress tensor. To assess the predictive capacity of different variants of the TBNN, we also introduce the root mean squared error, computed on the 6 independent components of $\boldsymbol \tau_s^*$: 
\be
\mathrm{RMSE} = \sqrt{\frac{1}{6 N_\mathrm{data}} \sum_{m=1}^{N_\mathrm{data}} \sum_{i=1}^{3}\sum_{j=i}^{3}(\hat \tau_{s,ij}^* - \tau_{s,ij}^*)^2}.
\label{eq:rmse}
\ee

Although the expression given by Eq. \eqref{eq:tbnn} is very general, simpler variants of the decomposition into basis tensors might yield results of similar accuracy. To that end, Table \ref{tab:TBNN_models_comparison} compares the RMSE of different versions of the TBNN model, where we vary both the number of tensors in the basis and the scalars fed to the network. We also introduce in {Table \ref{tab:TBNN_models_comparison}} the RMSE of the eddy-viscosity models studied in previous section for comparison. First, it comes out that, even though the 14-marker eddy-viscosity model was shown to be more accurate than the 3-marker model to predict the meso-scale viscosity, i.e. the norm of the stress tensor, both models have similar RMSE computed on the individual components of the deviatoric stress tensor. Thus, the improvement achieved by increasing the number of markers from four to fourteen is only marginal.  In contrast, all variants of the TBNN model show better accuracy. This highlights the limitation of the eddy-viscosity concept for subgrid stress modeling in gas-solid flows (just as in turbulent single-phase flows). A first TBNN model can be built only with tensors $\mathbf{T}^{(1)}$, $\mathbf{T}^{(3)}$ and the scalars $\lambda_1$ and $\lambda_3$, i.e. discarding quantities that involve the rotation-rate tensor $\Rs$. This model shows reduced RMSE with respect to the two eddy-viscosity models. However, models that account for the first four tensors in the basis display much better performance. Among these models, we observe that using only the first three scalars $\lambda_1$ to $\lambda_3$ defined by Eq. \eqref{eq:scalar_basis_tbnn} appear sufficient. The last two scalars $\lambda_4$ and $\lambda_5$ do not seem to provide extra information in order to capture the solid phase subgrid stress tensor. Likewise, the addition of the slip velocity vector as input marker only marginally reduces the RMSE value. Finally, using the complete tensor basis ($\mathbf T^{(1)}$ to $\mathbf T^{(10)}$) reduces only slightly the error made on the predicted stress tensor, while increasing the complexity of the network and the computational cost of the convergence algorithm. Therefore, for a practical use in fTFM simulations, we suggest to use the TBNN model with the first four basis tensors ($\mathbf T^{(1)}$ to $\mathbf T^{(4)}$), and, as an input to the network, the first three scalars of Pope's basis ($\lambda_1$ to $\lambda_3$), the filter size $\bar \Delta$ and the filtered solid volume fraction $\bar \phi_s$. 

The relative contribution of the different basis tensors in the construction of the final stress tensor can be estimated by the mean absolute value of the scalar functions $g^{(1)}$ to $g^{(4)}$. This analysis shows that the eddy-viscosity term, i.e. $g^{(1)}$, only contributes to 0.6\% of the final model, while the second, third and fourth tensors contribute respectively to 57.9\%, 24.6\% and 16.9\%. Future works should therefore investigate why the eddy-viscosity term virtually vanishes when a more complete tensor basis is used to build the deviatoric stress tensor and whether this conclusion is valid for a wide range of regimes.

\begin{figure}
    \centering
    \begin{subfigure}{0.48\textwidth}
    \includegraphics[width=\textwidth]{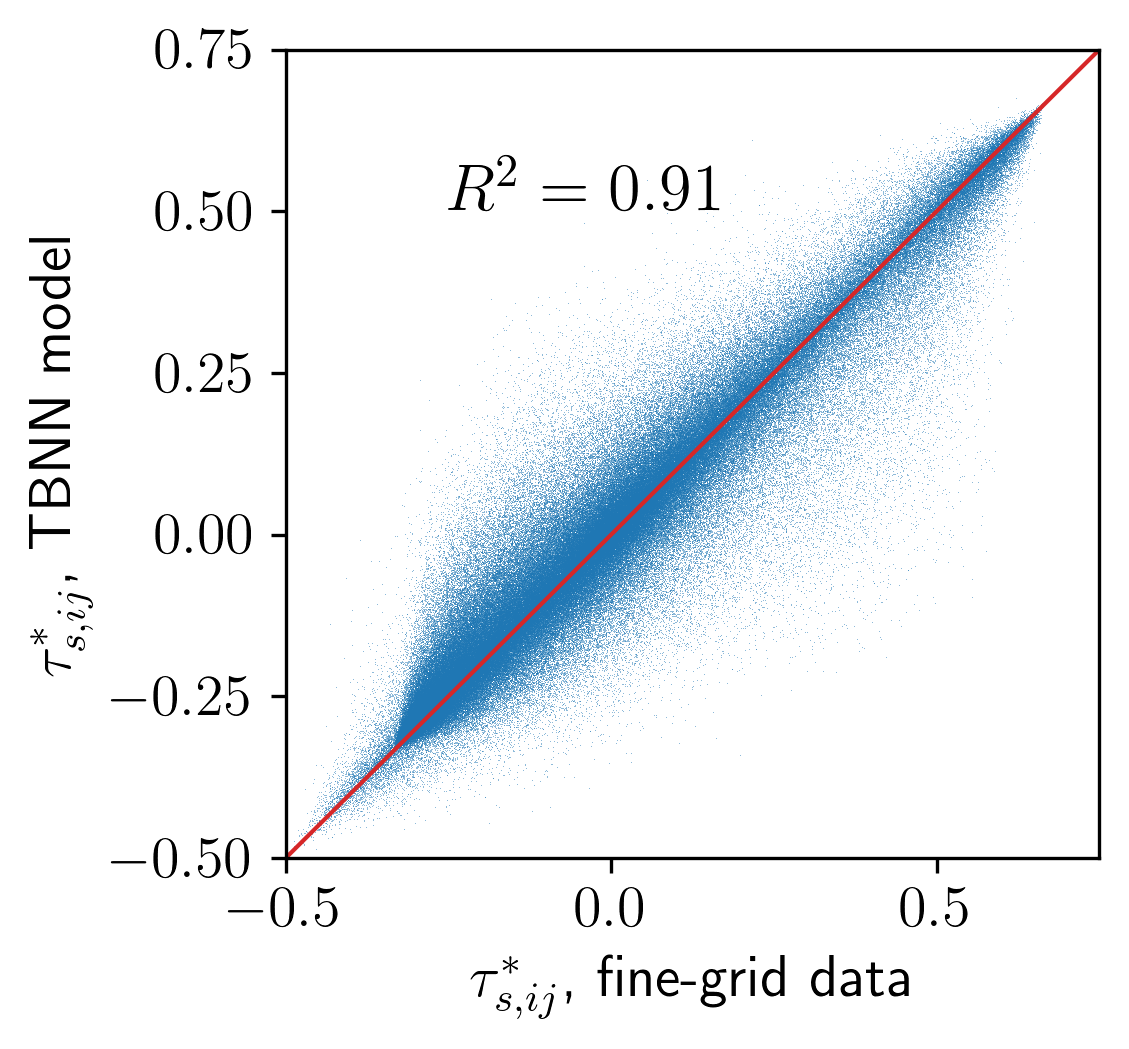}
     \end{subfigure}
    \hfill
    \begin{subfigure}{0.46\textwidth}
    \includegraphics[width=\textwidth]{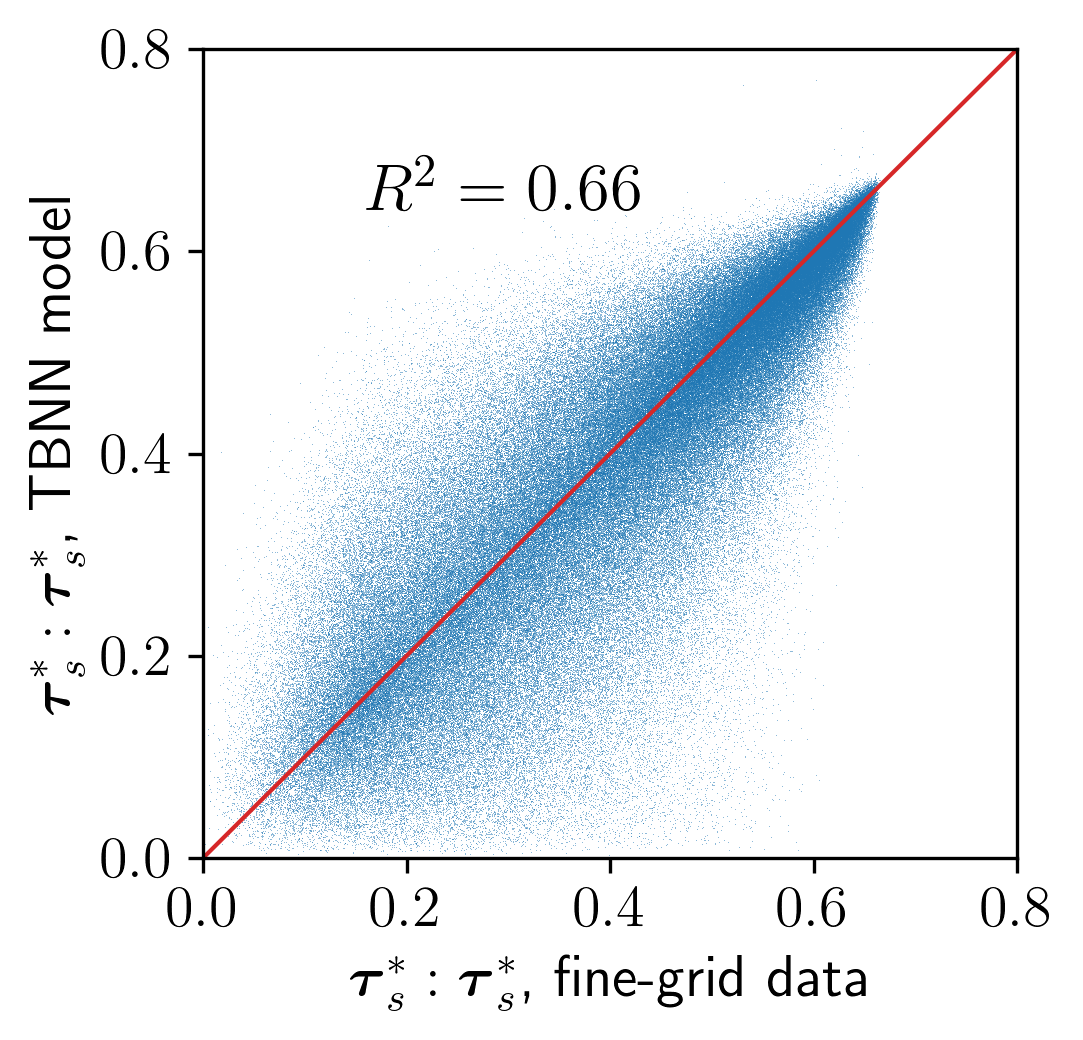}
     \end{subfigure}
     \caption{Prediction of the TBNN model built with four basis tensors ($\mathbf T^{(1)}$ to $\mathbf T^{(4)}$) and 5 input markers ($\lambda_1$, $\lambda_2$, $\lambda_3$, $\bar \phi_s$, $\bar \Delta$) for the 6 independent components of the solid phase deviatoric subgrid stress tensor (left) and the square of its norm (right) in Case 1.}
    \label{fig:TBNN_tau_predictions}
\end{figure}

Figure \ref{fig:TBNN_tau_predictions} shows the parity plots between the predicted values and the filtered fine-grid data for the 6 individual components of tensor $\boldsymbol \tau_s^*$ and for the square of its norm $\boldsymbol \tau_s^*:\boldsymbol \tau_s^*$. It appears that the scatter increases substantially in this second case. If the norm of the stress tensor is of higher importance for the prospected fTFM simulation, the loss function of the network could be customized for a more balanced error between the individual components and the norm of the tensor. 


\begin{table}[]
    \centering
    \begin{tabular}{l|l|l}
        Model & Markers &  RMSE \\
        \hline
        Eddy-viscosity ANN model &  $\lambda_1$, $\bar \phi_s$, $\bar \Delta$ & 0.4170 \\
        Eddy-viscosity ANN model & $\tilde \bu_s$, $\nabla \tilde \bu_s$, $\bar \phi_s$, $\bar \Delta$ & 0.4048 \\
        TBNN model - $\mathbf{T}^{(1)}$, $\mathbf{T}^{(3)}$ & $\lambda_1$, $\lambda_3$, $\bar \phi_s$, $\bar \Delta$ & 0.2076\\
        TBNN model - $\mathbf{T}^{(1)}$ to $\mathbf{T}^{(4)}$ & $\lambda_1$ to $\lambda_3$, $\bar \phi_s$, $\bar \Delta$ & 0.0697 \\
        TBNN model - $\mathbf{T}^{(1)}$ to $\mathbf{T}^{(4)}$ & $\lambda_1$ to $\lambda_3$, $\bar \phi_s$, $\tilde \bu_{slip}$, $\bar \Delta$ & 0.0695 \\
        TBNN model - $\mathbf{T}^{(1)}$ to $\mathbf{T}^{(4)}$ & $\lambda_1$ to $\lambda_5$, $\bar \phi_s$, $\bar \Delta$ & 0.0697 \\
        TBNN model - $\mathbf{T}^{(1)}$ to $\mathbf{T}^{(10)}$ & $\lambda_1$ to $\lambda_5$, $\bar \phi_s$, $\bar \Delta$ & 0.0692 \\
    \end{tabular}
    \caption{Comparison between linear eddy-viscosity and various TBNN models to predict the solid phase deviatoric subgrid stress tensor.}
    \label{tab:TBNN_models_comparison}
\end{table}


\section{Publicly Available ANN Python Library for fTFM Closures}
As we discussed before, the Python ML model source codes are available in the GitHub repository: \url{https://github.com/bahardy/fTFM\_ANN\_modeling.git}. Instructions for interested end-users who need to develop ANN models for the filtered drag force and the filtered solid phase stresses through TBNN using their own filtered datasets are as follows: The code reads the filtered dataset from a \textit{txt} file, which should be generated by the end-user by filtering fine-grid simulation results. As an illustrative example, and to avoid sharing a very large dataset of filtered simulation results generated by \textit{neptune\_cfd} ~\cite{Neau2020}, we have uploaded about 3\% of the filtered simulation results for a benchmark case. After training the networks, the output files from the training process are saved using Keras API \cite{chollet2021}: 1) a JSON file that contains the neural network structure; 2) a HDF5 file that contains weight information required for the evaluation of the neural network model. These files could then be read by an open-source interface Fortran/C++ code, which are suitable for MFIX and OpenFOAM simulation platforms. This implementation allows the user to read the flow quantities during simulation runtime and evaluate the prediction with ANN models. With ANN models and implementation approach, it is possible to assess its accuracy in \textit{a posteriori} simulations.  

\section{Conclusion}
\citet{Jiang2018,Jiang2021} proposed an artificial neural network (ANN) model for the dense flow regime in which the filter size, the filtered particle volume fraction, the filtered slip velocity, and the filtered gas phase pressure gradient, which are available in an fTFM model simulation are used to estimate the drift flux. The drift flux is then used to estimate the correction to the drag force for fTFM simulations, which are feasible for industrial-scale gas-solid flows. \citet{Rauchenzauner2022} used the filtered particle volume fraction, the scalar variance of the sub-grid particle volume fraction variation, and the kinetic energy associated with the subgrid velocity fluctuations of the particles to find the drift velocity in the dense flow regime. Both approaches work well when applied to a common dataset generated by filtering the results from a dense fluidized bed simulation. 
 
We then extended these studies through gas-particle fluidization simulations in periodic domains in the dilute regime and examined several different approaches to finding the correction to the drag force needed for fTFM models. It was found that the approach adopted by \citet{Jiang2021} for the dense flow regime works well for the dilute flow regime as well. This implies that a single ANN model that covers both regimes can be found by pooling together the dense and dilute flow regime datasets.
 
Furthermore, we introduce a Galilean-invariant tensor-based neural network (TBNN) model to capture the anisotropic particle phase stress stemming from the subgrid velocity fluctuations, which need a closure for fTFM approach. The proposed approach first utilizes distinct ANNs to find the filtered solid phase pressure and effective viscosity, which is a classical way of turbulence modeling in single-phase flows for the subgrid velocity fluctuations. It then employs a TBNN model to find the components of the filtered solid phase stress tensor. It is demonstrated that the TBNN approach captures the anisotropy quite nicely.
 
Future work should strive to generate a comprehensive drift flux model that combines the datasets generated through dense and dilute flow simulations. It should also examine how the level of sophistication of the stress model – a simple Smagorinsky-like eddy viscosity model vs. the TBNN model allowing for anisotropy – influences the predictions of fTFM simulations. A further step will be to perform fTFM simulations namely \textit{a posteriori} tests, with the developed models, and compare the predictions with the fine-grid TFM simulations and the experimental studies to assess their accuracy.

\section{Acknowledgements}
The \textit{netpune\_cfd} software is a multiphase CFD solver developed in the framework of the NEPTUNE project financially supported by EDF (Electricité de France), CEA (Commissariat à l'Energie Atomique), IRSN (Institut de Radioprotection et de Sureté Nucléaire) and Framatome. 

This work was financially supported by the Carnot ISIFoR Institute under the REBIS project (2022/2023) and by an Excellence Research Grant from Wallonie-Bruxelles International 2023/2024 (reference SOR/2023/550158). It also benefited from HPC resources provided by the supercomputing centers CALMIP (project P0111) and GENCI (allocation A0142B06012).

\bibliography{coarsegrained_models,ML}

\end{document}

%% file: symbols_and_cmd.tex
\def\AR{\text{\itshape\clipbox{0pt 0pt .28em 0pt}\AE\kern-.30emR}} 

\def\Fr{\operatorname{Fr}}

\newcommand{\br}{\mathbf{r}}

\def\Rey{\operatorname{Re}}

\def\u{\boldsymbol{u}}
\def\bu{\mathbf{u}}

\def\bv{\mathbf{v}}

\newcommand{\bx}{\mathbf{x}}

\newcommand{\by}{\mathbf{y}}

\newcommand{\beq}{\begin{equation}}
\newcommand{\beql}[1]{\begin{equation}\label{#1}}
\newcommand{\eeqc}{ \;\;,\end{equation}}
\newcommand{\eeq}{ \end{equation}}
\newcommand{\eeqd}{\;\;.\end{equation}}
\newcommand{\bseq}{\begin{subequations}}
\newcommand{\eseq}{\end{subequations}}

\newcommand{\eed}{.\end{equation}}
\newcommand{\eec}{,\end{equation}}

\newcommand{\bes}[1]{\begin{equation}\label{#1}\begin{split}}
\newcommand{\ees}{\end{split}\end{equation}}

\def\be{\begin{equation}}
\def\ee{\end{equation}}
\def\bea{\begin{eqnarray}}
\def\eea{\end{eqnarray}}
\def\bea{\begin{align}}
\def\ena{\end{align}}

\definecolor{darkGreen}{RGB}{0,160,0}
\definecolor{darkBlue}{RGB}{51,51,204}
\definecolor{darkRed}{RGB}{160,0,0}
\definecolor{orange}{RGB}{255,120,0}

\definecolor{blue1}{rgb}{0,0.45,0.74}
\definecolor{green1}{rgb}{0,0.5,0.0}
\definecolor{red1}{rgb}{0.8,0.0,0.0}

\definecolor{matlabBlue}{rgb}{0, 0.4470, 0.7410}
\definecolor{matlabOrange}{rgb}{0.8500, 0.3250, 0.0980}
\definecolor{matlabYellow}{rgb}{0.9290, 0.6940, 0.1250}
\definecolor{matlabPurple}{rgb}{0.4940, 0.1840, 0.5560}
\definecolor{matlabGreen}{rgb}{0.4660, 0.6740, 0.1880}
\definecolor{matlabRed}{rgb} {0.7765, 0.1569, 0.0902}
\definecolor{matlabBlack}{rgb}{0, 0, 0}

\definecolor{myGreen}{rgb}{0.0706, 0.5333, 0,1921}
\definecolor{keyBlue}{rgb}{0.0392, 0.2274, 0.4235}
\definecolor{keyBlueTrans}{rgb}{0.4274, 0.4980, 0.5764}
\definecolor{keyGreen}{rgb}{0.05490, 0.38039, 0.003921}
\definecolor{keyGreenTrans}{rgb}{0.7019, 0.8313, 0.7176}
\definecolor{keyOrange}{rgb}{0.98039, 0.4980, 0.03529}
\definecolor{keyGray}{rgb}{0.50196, 0.50196, 0.50196}

\definecolor{ellCD0p0}{rgb}{0.0,.45,.74}
\definecolor{ellCD0p03}{rgb}{0.28,.49,.39}
\definecolor{ellCD0p10}{rgb}{0.63,.44,.44}
\definecolor{bellCD0p03}{rgb}{0.89,.49,.00}

\definecolor{mutaxi}{rgb}{0.2,.9,1}
\definecolor{mulow}{rgb}{0.3,.75,.93}
\definecolor{muhigh}{rgb}{0.08,.17,.55}

\tikzset{cross/.style={cross out, draw, minimum size=2*(#1-\pgflinewidth), inner sep=0pt, outer sep=0pt}}

%% file: MLP_filteredDrag_drawing.tex

\tikzset{%
  every neuron/.style={
    circle,
    draw,
    minimum size=0.6cm
  },
  neuron missing/.style={
    draw=none, 
    scale=3,
    fill=none,
    text height=0.333cm,
    execute at begin node=\color{black}$\vdots$
  },
}

\begin{tikzpicture}[x=1.5cm, y=1.5cm, >=stealth]

\foreach \m/\l [count=\y] in {1,2,missing,3}
  \node [fill=blue, every neuron/.try, neuron \m/.try] (input-\m) at (0,2.5-\y) {};


\foreach \m [count=\y] in {1,2,3,missing,4}
  \node [fill=gray,every neuron/.try, neuron \m/.try ] (hidden1-\m) at (1.5,4-\y*1.4) {};
  
\foreach \m [count=\y] in {1,2,3,missing,4}
  \node [fill=gray, every neuron/.try, neuron \m/.try ] (hidden2-\m) at (3,4-\y*1.4) {};
  
\foreach \m [count=\y] in {1,2,3,missing,4}
  \node [fill=gray, every neuron/.try, neuron \m/.try ] (hidden3-\m) at (4.5,4-\y*1.4) {};


\foreach \m [count=\y] in {1}
  \node [fill=red, every neuron/.try, neuron \m/.try ] (output-\m) at (6,\y-1) {};

\foreach \l [count=\i] in {1,2,n}
  \draw [<-] (input-\i) -- ++(-1,0)
    node [above, midway] {$x_\l$};


\foreach \l [count=\i] in {1}
  \draw [->] (output-\i) -- ++(1,0)
    node [above, midway] { $y$};

\foreach \i in {1,...,3}
  \foreach \j in {1,...,4}
    \draw [->] (input-\i) -- (hidden1-\j);
    
 \foreach \i in {1,...,4}
  \foreach \j in {1,...,4}
    \draw [->] (hidden1-\i) -- (hidden2-\j);
    
 \foreach \i in {1,...,4}
  \foreach \j in {1,...,4}
    \draw [->] (hidden2-\i) -- (hidden3-\j);
    
\foreach \i in {1,...,4}
  \foreach \j in {1}
    \draw [->] (hidden3-\i) -- (output-\j);

\node [align=center, above] at (-0.5,2) {Input layer};

\node [align=center, above] at (3,3) {Hidden layers};

\node [align=center, above] at (6.5,0.8) {Output layer}; 

\end{tikzpicture}

%% file: TBNN_stresses_drawing.tex


\tikzset{%
  every neuron/.style={
    circle,
    draw,
    minimum size=0.6cm
  },
  neuron missing/.style={
    draw=none, 
    scale=3,
    fill=none,
    text height=0.333cm,
    execute at begin node=\color{black}$\vdots$
  },
}

\begin{tikzpicture}[x=1.5cm, y=1.5cm, >=stealth]

\foreach \m/\l [count=\y] in {1,2,missing,3}
  \node [fill=blue, every neuron/.try, neuron \m/.try] (input-\m) at (0,2.5-\y) {};

\foreach \m [count=\y] in {1,2,3,missing,4}
  \node [fill=gray,every neuron/.try, neuron \m/.try ] (hidden1-\m) at (1.5,4-\y*1.4) {};
  
\foreach \m [count=\y] in {1,2,3,missing,4}
  \node [fill=gray, every neuron/.try, neuron \m/.try ] (hidden2-\m) at (3,4-\y*1.4) {};
  
\foreach \m [count=\y] in {1,2,3,missing,4}
  \node [fill=gray, every neuron/.try, neuron \m/.try ] (hidden3-\m) at (4.5,4-\y*1.4) {};

\foreach \m [count=\y] in {1,2,missing,3}
  \node [fill=red, every neuron/.try, neuron \m/.try ] (output-\m) at (6,2.5-\y) {};


\foreach \m/\l [count=\y] in {1,2,missing,3}
  \node [fill=cyan, every neuron/.try, neuron \m/.try] (T_input-\m) at (6,-3-\y) {};

\foreach \m/\l [count=\y] in {1}
  \node [fill=violet, every neuron/.try, neuron \m/.try] (T_output-\m) at (8.5,-1-\y) {};
  
\foreach \l [count=\i] in {1,2,m}
  \draw [<-] (input-\i) -- ++(-1,0)
    node [above, midway] {$\lambda_\l$};

\foreach \l [count=\i] in {1,2,n}
  \draw [-] (output-\i) -- ++(1+0.2*\i,0)
    node [above, midway] { $g_\l$};

\foreach \l [count=\i] in {1,2,n}
  {
  \draw [<-] (T_input-\i) -- ++(-1,0)  node [above, near end] {$\mathbf T_\l$ \small{(3x3)}}; 
  \draw [-] (T_input-\i) -- ++(1+0.2*\i,0) -- ++(0,5.5); 
  \coordinate (X-\i) at  ($(T_input-\i) + (1+0.2*\i,0) + (0,2.75)$) ;
  }
\foreach \l [count=\i] in {1}
  \draw [->] (T_output-\i) -- ++(1+0.2*\i,0)
    node [above, near end] {$\mathbf a$ \small{(3x3)}};

\foreach \i in {1,...,3}
  \foreach \j in {1,...,4}
    \draw [->] (input-\i) -- (hidden1-\j);
    
 \foreach \i in {1,...,4}
  \foreach \j in {1,...,4}
    \draw [->] (hidden1-\i) -- (hidden2-\j);
    
 \foreach \i in {1,...,4}
  \foreach \j in {1,...,4}
    \draw [->] (hidden2-\i) -- (hidden3-\j);
    
\foreach \i in {1,...,4}
  \foreach \j in {1,...,3}
    \draw [->] (hidden3-\i) -- (output-\j);
    
\foreach \i in {1,...,3}    
 \foreach \j in {1}
    \draw [->] (X-\i) -- (T_output-\j);

\node [align=center, above] at (0,2) {Invariant \\ input layer};

\node [align=center, above] at (3,3) {Hidden layers};

\node [align=center, above] at (6,2) {Last hidden \\layer}; 

\node [align=center, above] at (6,-8.2) {Tensor input \\ layer}; 

\node [align=center, above] at (8.7,-1.2) {Merge output \\ layer}; 

\end{tikzpicture}
